\documentclass[english,twoside,11pt]{article}

\usepackage[german,english]{babel}
\usepackage{amsmath,amsthm,amssymb,latexsym}
\usepackage{moreverb}
\usepackage{url}
\usepackage{graphicx}
\usepackage{longtable}
\usepackage{lscape}
\usepackage{booktabs}
\usepackage{tabularx}
\usepackage{psfrag}
\usepackage{multicol}
\usepackage{paralist}
\usepackage{auto-pst-pdf}
\usepackage{color}

\newtheorem{deff}{Definition}
\newtheorem{lem}[deff]{Lemma}

\newtheorem{thm}[deff]{Theorem}
\newtheorem{cor}[deff]{Corollary}

\newtheorem{example}[deff]{Example}

\makeatletter

\title{Random Discrete Morse Theory \\and a New Library of Triangulations}

\author{\Large Bruno Benedetti\footnote{Supported by the Swedish Research Council, grant ``Triangulerade M{\aa}ngfalder, 
Knutteori i diskrete Morseteori'' and the DFG Collaborative Research Center TRR 109,
``Discretization in Geometry and Dynamics''.}\,\, and Frank H.~Lutz\footnote{Supported by the DFG Research Group ``Polyhedral Surfaces'', 
by \textsc{VILLUM FONDEN} through the Experimental Mathematics Network and by the Danish National Research Foundation (DNRF) through the Centre for Symmetry and Deformation.}}

\date{}

\begin{document}

\selectlanguage{english}

\maketitle

\enlargethispage*{3mm}

\begin{abstract}
(1) We introduce random discrete Morse theory as a computational scheme to measure the complicatedness of a triangulation. 
The idea is to try to quantify the frequence of  discrete Morse matchings with few critical cells. 
Our measure will depend on the topology of the space,  but also on how nicely the space is triangulated. 

(2) The scheme we propose looks for optimal discrete Morse functions with an elementary random heuristic.
Despite its na\"ivet\'e, this approach turns out to be very successful even in the case of huge inputs.

(3) In our view the existing libraries of examples in computational topology are `too easy' 
for testing algorithms based on discrete Morse theory.
We propose a new library containing more complicated (and thus more meaningful) test examples.
\end{abstract}

\section{Introduction}

Libraries of objects for algorithm testing are extremely common in computational geometry. Their set-up requires particular care: If a library consists of objects `too easy to understand', 
then basically any algorithm would score great on them, thus making it impossible for the researcher to appreciate the efficiency of the algorithms.
Of course, agreeing on what examples should be regarded as `easy' is a hard challenge, and how to quantify complicatedness is even harder. 

In the present paper, we focus on computational topology, which deals with simplicial complexes in an abstract manner, i.e., without prescribing a shape, a volume, or the dimension of a Euclidean space in which they embed. 
We present a possible random approach, which we call \emph{random discrete Morse theory}. 
The mathematical background 
relies on Forman's discrete Morse theory from 1998 \cite{Forman1998,Forman2002}, which in turn builds on 
Whitehead's simple homotopy theory, developed around 1939 \cite{Whitehead1939}. (Especially important is Whitehead's notion of collapsibility, which is a combinatorial strengthening of the contractibility property.)

\enlargethispage*{8mm}

Our idea is to create a quantitative version of these two theories. For example, we would like
to be able to tell not only if a complex is collapsible or not, but also `how easy it is' to find a collapsing sequence.
To give a mathematical basis to this intuition, we consider a random model where we perform elementary 
collapses completely at random. The probability  to find a complete collapsing sequence this way,
will measure how easy it is to collapse the complex. Although this probability is, in most cases,
too difficult to compute, we can estimate it empirically in polynomial time. 
The following elementary heuristic takes also into account complexes that are not contractible.

\bigskip

\noindent
\textsc{Algorithm: Random Discrete Morse}

\medskip

\noindent
\textsc{Input}: A $d$-dimensional (abstract, finite) simplicial complex $C$, given by its list of facets. 

\begin{compactitem}
\item[(0)] Initialize $c_0 = c_1 = \, \ldots \, = c_d =0$.
\item[(1)] Is the complex empty? If yes, then STOP; otherwise, go to (2).
\item[(2)] Are there free codimension-one faces? If yes, go to (3); if no, go to (4).
\item[(3)] \emph{Elementary Collapse}: Pick one free codimension-one face uniformly at random and delete it, together with
                 the unique face that contains it. Go back to (1).
\item[(4)] \emph{Critical Face}: Pick one of the top-dimensional faces uniformly at random and delete it from the complex. If $i$ is the dimension of the face just deleted, increment $c_i$ by 1 unit. Go back to (1).
\end{compactitem}

\noindent
\textsc{Output}: The resulting discrete Morse vector $(c_0, c_1, c_2, \ldots, c_d)$. 

\bigskip

By construction, $c_i$ counts the critical faces of dimension $i$. 
(We do not consider the empty face as a free face.)
According to Forman~\cite{Forman1998}, 
any discrete Morse vector $(c_0, c_1, c_2, \ldots, c_d)$ is also the face vector of a cell complex homotopy equivalent to $C$.

 \begin{deff}
The \emph{discrete Morse spectrum} $\sigma$ of\, a (finite) simplicial complex\, $C$ is the collection of all possible resulting discrete Morse vectors produced by the algorithm
 \textsc{Random Discrete Morse}  
 together with the distribution of the respective probabilities.
 \end{deff}

\begin{figure}[t]
\begin{center}
\includegraphics[width=3.5cm]{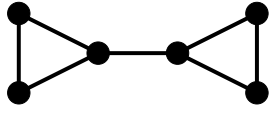}
\end{center}
\caption{The graph $A_7$.}
\label{fig:A7}
\end{figure}

\emph{Example}: Consider the graph $A_7$ of Figure~\ref{fig:A7} above. As there are no free vertices in it, \textsc{Random Discrete Morse} picks an edge uniformly at random and deletes it. If the edge chosen is the central bridge (which happens with probability  $\frac{1}{7}$), the output discrete Morse vector is $(2,3)$. If any other edge than the central one is chosen, the output vector is $(1,2)$. The discrete Morse spectrum is therefore $\{\mbox{$\frac{6}{7}$-$(1,2)$, $\frac{1}{7}$-$(2,3)$}\}$;
or, shortly, $\{(1,2), (2,3)\}$ if we simply want to list the vectors of the spectrum. 
                               
\bigskip

The algorithm \textsc{Random Discrete Morse} requires no backtracking, and `digests' the complex very rapidly. The output $(1,0,0, \ldots, 0)$ is a certificate of collapsibility.
If the output is different from $(1,0, \ldots, 0)$, the complex could still be collapsible with a different sequence of free-face deletions. 
 \textsc{Random Discrete Morse} declares a $k$-face critical only if there are no free $(k-1)$-faces available. This keeps the number of faces declared critical to a minimum, thus making it more likely for the output vector to be optimal. Unfortunately, there are complexes on which the probability to achieve the optimal discrete Morse vector can be arbitrarily small;
 see Appendix A and also \cite{AdiprasitoBenedettiLutz2013pre} for a further discussion.  But in case optimality is not reached, the algorithm still outputs something meaningful, namely (as already mentioned) the $f$-vector of a cell complex homotopy equivalent to the given complex. 

Since the output arrives quickly, we can re-launch the program, say, 10000 times, possibly on separate computers (independently). 
The distribution of the obtained outcomes yields an approximation of the discrete Morse spectrum.
By the so-called Morse inequalities, each output vector is componentwise larger or equal 
than the vector of Betti numbers $(\beta_0, \ldots, \beta_d)$. When the spectrum stays `close' to the vector of Betti numbers, we can regard the triangulation to be easy. This allows an empirical analysis of how complicated the complex is.

We point out that the problem of finding \emph{optimal} discrete Morse functions (with as fewest critical cells as possible)
is {\cal NP}-hard \cite{JoswigPfetsch2006,LewinerLopesTavares2003a}; even the decision problem
of whether some given (connected, finite) simplicial complex is collapsible or not is {\cal NP}-complete \cite{Tancer2012pre}.
We therefore \emph{should not} expect to immediately find optimal discrete Morse vectors for any input. 
Indeed, one can easily construct examples on which our (or similar) random heuristic performs poorly; see Appendix A.

However, for many triangulations even of \emph{huge} size,  our elementary random heuristic produces optimal discrete Morse functions 
in almost 100\% of the runs of the program. 
This could be interesting in the future also for homology computations.
Discrete Morse functions (for general cell complexes) are implicitly computed in several homology algorithms 
that are based on fast (co-)reduction techniques, like the packages CHomP~\cite{Chomp}, RedHom~\cite{RedHom},
and Perseus~\cite{Perseus}.

\enlargethispage*{2mm}

The paper is structured as follows: First we give details of our algorithm (Section~\ref{sec:details}) and 
compare it with previous approaches (Section~\ref{sec:comparison}). Then we survey the existing topological 
and combinatorial lower bounds for optimal discrete Morse vectors (Section~\ref{sec:lower bounds}). 
Finally, we describe and examine a collection of examples coming from several different areas 
of topology (Section~\ref{sec:computational_results}).
In our opinion, the resulting library (Appendix B) is a richer and more sensitive testing ground for implementations
based on discrete Morse theory.

\section{Details of the algorithm and computational complexity}
\label{sec:details}

\begin{figure}[t]
\begin{center}
\begin{postscript}
\psfrag{1}{1}
\psfrag{2}{2}
\psfrag{3}{3}
\psfrag{4}{4}
\psfrag{5}{5}
\includegraphics[width=3.3cm]{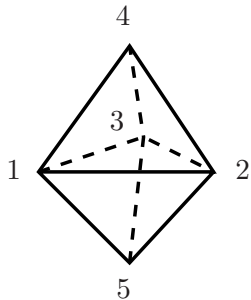}
\end{postscript}
\end{center}
\caption{The bipyramid.}
\label{fig:bipyramid}
\end{figure}

In the following, we give a more explicit description of our random heuristic.

The first thing we do is to build the Hasse diagram of the given simplicial complex $C$, 
which represents the incidence structure of the face poset of $C$; see Figure~\ref{fig:hasse_bipyramid} 
for an example.
It takes $O(d\cdot I\cdot T)$ steps to construct the Hasse diagram
of a simplicial complex, in case the complex is given by its
facets (or to be precise, by its vertex-facet incidences). 
Here $d$ is the dimension of the input complex, $T$ the total number
of faces, and $I$ the number of vertex-facet-incidences 
\cite{KaibelPfetsch2002}; cf.\ also \cite{KaibelPfetsch2003}.

\enlargethispage*{4mm}

Once the upward Hasse diagram and the downward Hasse diagram are set up (see below),
we deconstruct a copy of the upward Hasse diagram in every run of our program by deleting (randomly picked) critical faces or pairs
of faces in case there are free faces. We illustrate this with a concrete example. 

\begin{example}[The bipyramid] \rm
The $2$-dimensional boundary of the bipyramid of Figure~\ref{fig:bipyramid} has $6$~triangles, $9$~edges, and $5$~vertices.
We list the faces level-wise in lexicographic order and identify
each face by a label $k^i$ denoting the $k$-th face of dimension $i$
in the respective lexicographic list.

\begin{center}
$\emph{1}^{\emph{\,2}}$: \mbox{[1,2,4]},\, $\emph{2}^{\emph{\,2}}$: \mbox{[1,2,5]},\, $\emph{3}^{\emph{\,2}}$: \mbox{[1,3,4]},\, 
$\emph{4}^{\emph{\,2}}$: \mbox{[1,3,5]},\, $\emph{5}^{\emph{\,2}}$: \mbox{[2,3,4]},\, $\emph{6}^{\emph{\,2}}$: \mbox{[2,3,5]}, \\[2mm]
$\emph{1}^{\emph{\,1}}$: \mbox{[1,2]},\, $\emph{2}^{\emph{\,1}}$: \mbox{[1,3]},\, $\emph{3}^{\emph{\,1}}$: \mbox{[1,4]},\, 
$\emph{4}^{\emph{\,1}}$: \mbox{[1,5]},\, $\emph{5}^{\emph{\,1}}$: \mbox{[2,3]},\, $\emph{6}^{\emph{\,1}}$: \mbox{[2,4]},\, 
$\emph{7}^{\emph{\,1}}$: \mbox{[2,5]},\, $\emph{8}^{\emph{\,1}}$: \mbox{[3,4]},\, $\emph{9}^{\emph{\,1}}$: \mbox{[3,5]}, \\[2mm]
$\emph{1}^{\emph{\,0}}$: \mbox{[1]},\, $\emph{2}^{\emph{\,0}}$: \mbox{[2]},\, $\emph{3}^{\emph{\,0}}$: \mbox{[3]},\, 
$\emph{4}^{\emph{\,0}}$: \mbox{[4]},\, $\emph{5}^{\emph{\,0}}$: \mbox{[5]}. 
\end{center}

Next, we initialize the Hasse diagram. Hereby, the graph of the Hasse diagram is stored twice.
In the \emph{upward Hasse diagram}, we level-wise list all inclusions of $i$-dimensional faces 
in $(i+1)$-dimen\-sional faces,

\medskip

\noindent
$i=1$:\, $\emph{1}^{\emph{\,1}}\!\nearrow\{\emph{1}^{\emph{\,2}},\emph{2}^{\emph{\,2}}\}$,\, 
         $\emph{2}^{\emph{\,1}}\!\nearrow\{\emph{3}^{\emph{\,2}},\emph{4}^{\emph{\,2}}\}$,\,
         $\emph{3}^{\emph{\,1}}\!\nearrow\{\emph{1}^{\emph{\,2}},\emph{3}^{\emph{\,2}}\}$,\,
         $\emph{4}^{\emph{\,1}}\!\nearrow\{\emph{2}^{\emph{\,2}},\emph{4}^{\emph{\,2}}\}$,\,
         $\emph{5}^{\emph{\,1}}\!\nearrow\{\emph{5}^{\emph{\,2}},\emph{6}^{\emph{\,2}}\}$, \\
\mbox{}\hspace{10mm}
         $\emph{6}^{\emph{\,1}}\!\nearrow\{\emph{1}^{\emph{\,2}},\emph{5}^{\emph{\,2}}\}$,\,
         $\emph{7}^{\emph{\,1}}\!\nearrow\{\emph{2}^{\emph{\,2}},\emph{6}^{\emph{\,2}}\}$,\,
         $\emph{8}^{\emph{\,1}}\!\nearrow\{\emph{3}^{\emph{\,2}},\emph{5}^{\emph{\,2}}\}$,\,
         $\emph{9}^{\emph{\,1}}\!\nearrow\{\emph{4}^{\emph{\,2}},\emph{6}^{\emph{\,2}}\}$, \\[2mm]
$i=0$:\, $\emph{1}^{\emph{\,0}}\!\nearrow\{\emph{1}^{\emph{\,1}},\emph{2}^{\emph{\,1}},\emph{3}^{\emph{\,1}},\emph{4}^{\emph{\,1}}\}$,\,
         $\emph{2}^{\emph{\,0}}\!\nearrow\{\emph{1}^{\emph{\,1}},\emph{5}^{\emph{\,1}},\emph{6}^{\emph{\,1}},\emph{7}^{\emph{\,1}}\}$,\,
         $\emph{3}^{\emph{\,0}}\!\nearrow\{\emph{2}^{\emph{\,1}},\emph{5}^{\emph{\,1}},\emph{8}^{\emph{\,1}},\emph{9}^{\emph{\,1}}\}$, \\
\mbox{}\hspace{10mm}
         $\emph{4}^{\emph{\,0}}\!\nearrow\{\emph{3}^{\emph{\,1}},\emph{6}^{\emph{\,1}},\emph{8}^{\emph{\,1}}\}$,\,
         $\emph{5}^{\emph{\,0}}\!\nearrow\{\emph{4}^{\emph{\,1}},\emph{7}^{\emph{\,1}},\emph{9}^{\emph{\,1}}\}$,

\medskip

\noindent
while in the \emph{downward Hasse diagram} we level-wise list the $(j-1)$-dimensional faces that are contained
in the $j$-dimensional faces,

\medskip

\noindent
$j=2$:\, $\emph{1}^{\emph{\,2}}\!\searrow\{\emph{1}^{\emph{\,1}},\emph{3}^{\emph{\,1}},\emph{6}^{\emph{\,1}}\}$,\, 
         $\emph{2}^{\emph{\,2}}\!\searrow\{\emph{1}^{\emph{\,1}},\emph{4}^{\emph{\,1}},\emph{7}^{\emph{\,1}}\}$,\,
         $\emph{3}^{\emph{\,2}}\!\searrow\{\emph{2}^{\emph{\,1}},\emph{3}^{\emph{\,1}},\emph{8}^{\emph{\,1}}\}$,\\
\mbox{}\hspace{10mm}
         $\emph{4}^{\emph{\,2}}\!\searrow\{\emph{2}^{\emph{\,1}},\emph{4}^{\emph{\,1}},\emph{9}^{\emph{\,1}}\}$,\,
         $\emph{5}^{\emph{\,2}}\!\searrow\{\emph{5}^{\emph{\,1}},\emph{6}^{\emph{\,1}},\emph{8}^{\emph{\,1}}\}$,\,
         $\emph{6}^{\emph{\,2}}\!\searrow\{\emph{5}^{\emph{\,1}},\emph{7}^{\emph{\,1}},\emph{9}^{\emph{\,1}}\}$, \\[2mm]
$j=1$:\, $\emph{1}^{\emph{\,1}}\!\searrow\{\emph{1}^{\emph{\,0}},\emph{2}^{\emph{\,0}}\}$,\,
         $\emph{2}^{\emph{\,1}}\!\searrow\{\emph{1}^{\emph{\,0}},\emph{3}^{\emph{\,0}}\}$,\,
         $\emph{3}^{\emph{\,1}}\!\searrow\{\emph{1}^{\emph{\,0}},\emph{4}^{\emph{\,0}}\}$,\,
         $\emph{4}^{\emph{\,1}}\!\searrow\{\emph{1}^{\emph{\,0}},\emph{5}^{\emph{\,0}}\}$,\,
         $\emph{5}^{\emph{\,1}}\!\searrow\{\emph{2}^{\emph{\,0}},\emph{3}^{\emph{\,0}}\}$, \\
\mbox{}\hspace{10mm}
         $\emph{6}^{\emph{\,1}}\!\searrow\{\emph{2}^{\emph{\,0}},\emph{4}^{\emph{\,0}}\}$,\,
         $\emph{7}^{\emph{\,1}}\!\searrow\{\emph{2}^{\emph{\,0}},\emph{5}^{\emph{\,0}}\}$,\,
         $\emph{8}^{\emph{\,1}}\!\searrow\{\emph{3}^{\emph{\,0}},\emph{4}^{\emph{\,0}}\}$,\,
         $\emph{9}^{\emph{\,1}}\!\searrow\{\emph{3}^{\emph{\,0}},\emph{5}^{\emph{\,0}}\}$.\\

\noindent
Here, $\emph{3}^{\emph{\,1}}\!\nearrow\{\emph{1}^{\emph{\,2}},\emph{3}^{\emph{\,2}}\}$
is the short notation for the inclusion of the edge $\emph{3}^{\emph{\,1}}$:~\mbox{[1,4]}
in the two triangles $\emph{1}^{\emph{\,2}}$:~\mbox{[1,2,4]} and $\emph{3}^{\emph{\,2}}$:~\mbox{[1,3,4]}.

\begin{figure}
\begin{center}
\begin{postscript}
\psfrag{1}{1}
\psfrag{2}{2}
\psfrag{3}{3}
\psfrag{4}{4}
\psfrag{5}{5}
\psfrag{12}{12}
\psfrag{13}{13}
\psfrag{14}{14}
\psfrag{15}{15}
\psfrag{23}{23}
\psfrag{24}{24}
\psfrag{25}{25}
\psfrag{34}{34}
\psfrag{35}{35}
\psfrag{124}{124}
\psfrag{125}{125}
\psfrag{134}{134}
\psfrag{135}{135}
\psfrag{234}{234}
\psfrag{235}{235}
\psfrag{e}{$\emptyset$}
\includegraphics[width=15cm]{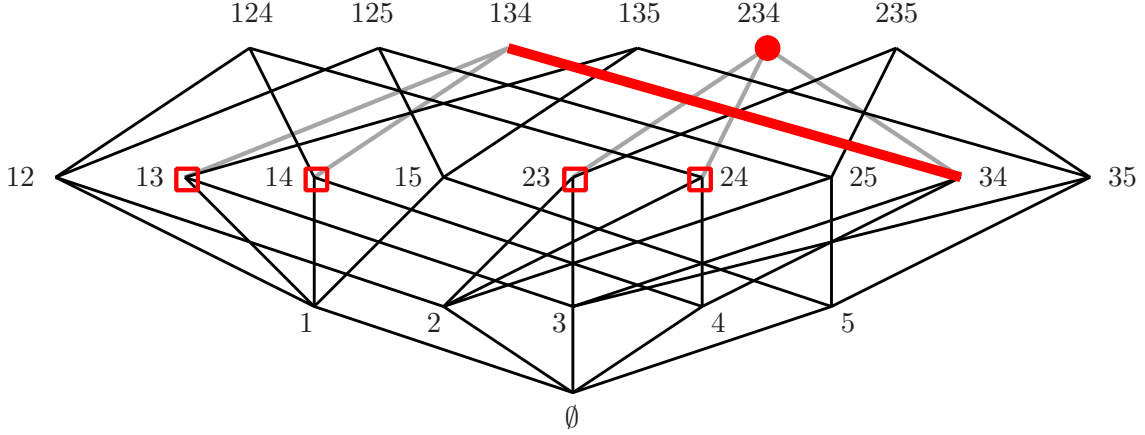}
\end{postscript}
\end{center}
\caption{The Hasse diagram of the bipyramid with one critical triangle (234), one matching edge (34--134), 
and four free edges (13, 14, 23, 24) highlighted.}
\label{fig:hasse_bipyramid}
\end{figure}

During each run, the downward Hasse diagram is maintained, while a
copy of the upward Hasse diagram
is updated after the removal of a critical face or of a pair consisting of a free face
and the unique face it is contained in. The sequence of updating steps for the above example
could be as follows:

\medskip

\noindent
0. \texttt{compute downward and upward Hasse diagram}\\
1. \texttt{initialize copy of the upward Hasse diagram}\\
2. \texttt{free edges}:\, none\\
3. \texttt{select random critical triangle:}\, $\emph{5}^{\emph{\,2}}$: \mbox{[2,3,4]} \\
4. \texttt{update upward Hasse diagram:}\\[2mm]
\mbox{}\hspace{5mm}
$i=1$:\, $\emph{1}^{\emph{\,1}}\!\nearrow\{\emph{1}^{\emph{\,2}},\emph{2}^{\emph{\,2}}\}$,\, 
         $\emph{2}^{\emph{\,1}}\!\nearrow\{\emph{3}^{\emph{\,2}},\emph{4}^{\emph{\,2}}\}$,\,
         $\emph{3}^{\emph{\,1}}\!\nearrow\{\emph{1}^{\emph{\,2}},\emph{3}^{\emph{\,2}}\}$,\,
         $\emph{4}^{\emph{\,1}}\!\nearrow\{\emph{2}^{\emph{\,2}},\emph{4}^{\emph{\,2}}\}$,\,
         $\emph{5}^{\emph{\,1}}\!\nearrow\{\emph{6}^{\emph{\,2}}\}$, \\
\mbox{}\hspace{16mm}
         $\emph{6}^{\emph{\,1}}\!\nearrow\{\emph{1}^{\emph{\,2}}\}$,\,
         $\emph{7}^{\emph{\,1}}\!\nearrow\{\emph{2}^{\emph{\,2}},\emph{6}^{\emph{\,2}}\}$,\,
         $\emph{8}^{\emph{\,1}}\!\nearrow\{\emph{3}^{\emph{\,2}}\}$,\,
         $\emph{9}^{\emph{\,1}}\!\nearrow\{\emph{4}^{\emph{\,2}},\emph{6}^{\emph{\,2}}\}$ \\[2mm]
5. \texttt{free edges:\, $\emph{5}^{\emph{\,1}},\emph{6}^{\emph{\,1}},\emph{8}^{\emph{\,1}}$}\\
6. \texttt{select random free edge:}\, $\emph{8}^{\emph{\,1}}$: \mbox{[3,4]}\, \texttt{paired with}\, $\emph{3}^{\emph{\,2}}$: \mbox{[1,3,4]} \\
7. \texttt{update upward Hasse diagram:}\\[2mm]
\mbox{}\hspace{5mm}
$i=1$:\, $\emph{1}^{\emph{\,1}}\!\nearrow\{\emph{1}^{\emph{\,2}},\emph{2}^{\emph{\,2}}\}$,\, 
         $\emph{2}^{\emph{\,1}}\!\nearrow\{\emph{4}^{\emph{\,2}}\}$,\,
         $\emph{3}^{\emph{\,1}}\!\nearrow\{\emph{1}^{\emph{\,2}},\emph{3}^{\emph{\,2}}\}$,\,
         $\emph{4}^{\emph{\,1}}\!\nearrow\{\emph{2}^{\emph{\,2}},\emph{4}^{\emph{\,2}}\}$,\,
         $\emph{5}^{\emph{\,1}}\!\nearrow\{\emph{6}^{\emph{\,2}}\}$, \\
\mbox{}\hspace{16mm}
         $\emph{6}^{\emph{\,1}}\!\nearrow\{\emph{1}^{\emph{\,2}}\}$,\,
         $\emph{7}^{\emph{\,1}}\!\nearrow\{\emph{2}^{\emph{\,2}},\emph{6}^{\emph{\,2}}\}$,\,
         $\emph{8}^{\emph{\,1}}\!\nearrow\{\}$,\,
         $\emph{9}^{\emph{\,1}}\!\nearrow\{\emph{4}^{\emph{\,2}},\emph{6}^{\emph{\,2}}\}$, \\[2mm]
\mbox{}\hspace{5mm}
$i=0$:\, $\emph{1}^{\emph{\,0}}\!\nearrow\{\emph{1}^{\emph{\,1}},\emph{2}^{\emph{\,1}},\emph{3}^{\emph{\,1}},\emph{4}^{\emph{\,1}}\}$,\,
         $\emph{2}^{\emph{\,0}}\!\nearrow\{\emph{1}^{\emph{\,1}},\emph{5}^{\emph{\,1}},\emph{6}^{\emph{\,1}},\emph{7}^{\emph{\,1}}\}$,\,
         $\emph{3}^{\emph{\,0}}\!\nearrow\{\emph{2}^{\emph{\,1}},\emph{5}^{\emph{\,1}},\emph{9}^{\emph{\,1}}\}$, \\
\mbox{}\hspace{16mm}
         $\emph{4}^{\emph{\,0}}\!\nearrow\{\emph{3}^{\emph{\,1}},\emph{6}^{\emph{\,1}}\}$,\,
         $\emph{5}^{\emph{\,0}}\!\nearrow\{\emph{4}^{\emph{\,1}},\emph{7}^{\emph{\,1}},\emph{9}^{\emph{\,1}}\}$\\[2mm]
8. \texttt{free edges:\, $\emph{2}^{\emph{\,1}},\emph{3}^{\emph{\,1}},\emph{5}^{\emph{\,1}},\emph{6}^{\emph{\,1}}$}\\
9. \texttt{\dots}

\medskip

The downward Hasse diagram tells us precisely which parts of the upward Hasse diagram we have to update.
For example, the choice of the critical triangle $\emph{5}^{\emph{\,2}}$: \mbox{[2,3,4]} 
forces us to update, via $\emph{5}^{\emph{\,2}}\!\searrow\{\emph{5}^{\emph{\,1}},\emph{6}^{\emph{\,1}},\emph{8}^{\emph{\,1}}\}$,
the inclusions of the edges $\emph{5}^{\emph{\,1}},\emph{6}^{\emph{\,1}},\emph{8}^{\emph{\,1}}$
in the upward Hasse diagram (by removing the triangle $\emph{5}^{\emph{\,2}}$ as including face). 
\end{example}

Triangulations of closed manifolds initially have no free faces. Thus, we start 
with an empty list of free faces and immediately remove a random critical face.
For triangulations of manifolds with boundary or general simplicial complexes,
we first have to initialize the list of free faces. (This extra effort in computation time
can be seen by comparing the respective run times for the examples \texttt{knot} and \texttt{nc\_sphere} 
in Table~\ref{tbl:discrete_morse_spectra} (see also Section~\ref{sec:complicated_balls_and_spheres}): 
With 0.813 seconds, the $3$-ball \texttt{knot} takes slightly
longer per round than the $3$-sphere \texttt{nc\_sphere} with 0.470 seconds.)

Whenever we are done with one level of the Hasse diagram, we initialize the set of free faces for the next level below.
Besides updating the upward Hasse diagram in each round, we also keep track of 
\begin{compactitem}
\item the current \emph{list of free faces} (and update this list whenever we delete a critical face or a pair 
consisting of a free face and the unique face it is contained in),
\item the current \emph{discrete Morse vector} $(c_0,c_1,\dots,c_d)$
  (which is initialized by $(0,0,\dots,0)$ and updated by incrementing
  $c_i$ by one whenever a critical face of dimension $i$ is selected).
\end{compactitem}
At the end of every round, the resulting discrete Morse vector $(c_0,c_1,\dots,c_d)$
is stored along with its number of appearances in the various rounds.
Eventually, we output the list of all obtained discrete Morse vectors
together with their frequencies.

\subsection{Implementation in GAP}

We implemented our random heuristic in GAP~\cite{GAP4}. 
In particular, we used GAP operations on lists and sets
to initialize and update Hasse diagrams and respective lists 
of free faces. Our implementation is basic and has roughly 
150 lines of code.

\enlargethispage*{5mm}

The largest complex (in terms of number of faces) we tested our program on has 
face vector $f=(5013,72300,290944,495912,383136,110880)$.
For this triangulation \cite{AdiprasitoBenedettiLutz2013pre} of a contractible $5$-manifold different (!) from a $5$-ball,
it took  in total 60:17:33\,+\,21:41:31 h:min:sec to first build
the Hasse diagram and then to run the random heuristic once. 
As resulting discrete Morse vector we obtained $(1,0,0,0,0,0)$; thus, this non-trivial $5$-manifold is collapsible \cite{AdiprasitoBenedettiLutz2013pre}.

We point out that there is considerable room for improvement with respect to computation time. 
First of all, the Hasse diagram of a complex can be stored 
in terms of (sparse) boundary matrices on which fast elimination steps 
represent elementary collapses; see Joswig~\cite{Joswig2004pre}
for a discussion. In addition, it is way faster to perform
matrix operations in, say, C++ compared to elementary set operations in GAP.
However, if it comes to compute (respectively to simplify a
presentation of) the fundamental group
of a simplicial complex, then GAP provides efficient heuristics;
cf.\ Section~\ref{sec:fundamental_groups}.

\section{Comparison with other algorithms}
\label{sec:comparison}

There are three main previous algorithmic approaches that aim to compute optimal discrete Morse functions 
for simplicial complexes, one by Joswig and Pfetsch~\cite{JoswigPfetsch2006},
one by Eng\-str\"om~\cite{Engstroem2009b}, and one by Lewiner, Lopes, and Tavares~\cite{LewinerLopesTavares2003a}
(cf.\ also Lewiner~\cite{Lewiner2005}). 

Tools that allow to improve discrete Morse functions were provided by Hersh~\cite{Hersh2005}.
King, Knudson, and Mramor \cite{HKingKnudsonMramor2005} discussed improving
discrete Morse functions for geometric complexes in ${\mathbb R}^3$.

A completely different random approach to discrete Morse theory was attempted already by Nicolaescu \cite{Nicolaescu2012pre}.
Essentially he tried to randomly choose edges in the Hasse diagram to obtain discrete Morse matchings, 
but showed that this approach will not be successful. Indeed, choosing  edges in the Hasse diagram at random
will produce bottlenecks even for complexes that are easily collapsible.

\subsection{The algorithm of Joswig and Pfetsch}

This deterministic algorithm, apart from a complete backtrack search, is currently the only available implementation 
that actually determines optimal discrete Morse functions for all inputs. 
In the Joswig--Pfetsch approach~\cite{JoswigPfetsch2006}, the problem of finding
an optimal discrete Morse function is translated into a maximal matching problem 
for the underlying graph of the Hasse diagram with an additional acyclicity condition \cite{Chari2000,Forman1998}. 
The acyclic matching problem is then solved as an integer linear program. 

\enlargethispage*{4mm}

For various small instances, the Joswig--Pfetsch approach successfully produces
optimal discrete Morse functions~\cite{JoswigPfetsch2006}.
A first case, however, for which the associated integer linear program 
was too large to handle is for the $16$-vertex triangulation \texttt{poincare}~\cite{BjoernerLutz2000} 
of the Poincar\'e homology $3$-sphere with $f=(16,106,180,90)$.
Joswig and Pfetsch interrupted the computation after one week (perhaps because they did not make use of the 
fact that at least six critical cells are necessary since the fundamental group 
of the Poincar\'e homology $3$-sphere is non-cyclic; cf.~Section~\ref{sec:classical}). 
For the same instance our heuristic found the optimal Morse vector $(1,2,2,1)$ within 0.02 seconds.
Also for other small instances our heuristic was much faster,
e.g., for Rudin's ball \texttt{rudin} \cite{Rudin1958,Wotzlaw2005} with $f=(14,66,94,41)$,
Joswig and Pfetsch needed 103.78 seconds to achieve the optimal discrete Morse vector $(1,0,0,0)$ 
while our heuristic found the optimum in 0.004+0.00107 seconds; cf.~Section~\ref{sec:complicated_balls_and_spheres}.

\subsection{The approach by Engstr\"om}

The heuristic approach by Engstr\"om~\cite{Engstroem2009b} is elegant and fast. Roughly speaking, the idea is to proceed by deleting vertex stars, rather than by deleting pairs of faces. 
Engstr\"om introduces what he calls `Fourier--Morse theory', a theory based on Kahn--Saks--Sturtevant's notion of non-evasiveness, much like Forman's discrete Morse theory was based on Whitehead's notion of collapsibility. 
Instead of computing discrete Morse functions, Engstr\"om's heuristic computes Fourier--Morse functions, which are \emph{some} discrete Morse functions, but not necessarily the optimal ones among them.
In particular, obtaining an output $(1,0,\ldots,0)$ with this approach yields a certificate of non-evasiveness, 
a stronger property than collapsibility; cf.~\cite{KahnSaksSturtevant1984}. 

\enlargethispage*{1mm}

However, there is a $3$-ball with only $12$ vertices which has the collapsibility property, 
but not the non-evasiveness property~\cite{BenedettiLutz2013apre}. 
As for other examples, Engstr\"om obtains $(1,5,5,1)$  as a discrete Fourier--Morse vector 
for the $16$-vertex triangulation of the Poincar\'e homology $3$-sphere \texttt{poincare}.
Instead, the optimal discrete Morse vector for this example is $(1,2,2,1)$. For Rudin's ball \texttt{rudin}
Engstr\"om found $(1,2,2,0)$ compared to the optimum $(1,0,0,0)$. 
Engstr\"om's implementation depends on the vertex-labeling of a complex. 
For a fixed labeling, the optimal discrete Morse vector is often missed, even on triangulations of relatively small size.

\subsection{The heuristic of Lewiner, Lopes, and Tavares}

The heuristic approach of Lewiner, Lopes, and Tavares~\cite{LewinerLopesTavares2003a}
(cf.\ also Lewiner~\cite{Lewiner2005}) is fast and was used to produce optimal discrete Morse vectors 
for several large $2$- and $3$-dimensional complexes.
The problem of finding optimal discrete Morse vectors is reformulated in terms
of finding maximal hyperforests of hypergraphs. Then different greedy heuristics
are used to obtain large hyperforests.

It has to be remarked, though, that most of the instances listed in \cite{LewinerLopesTavares2003a}
and later in~\cite{Lewiner2005} are mostly harmless from the point of view of discrete Morse theory;
they mainly are $2$-dimensional surfaces or shellable $3$-dimensional balls and
spheres, or products thereof --- with the exception of
the three more complicated examples \texttt{knot}, \texttt{nc\_sphere}, and \texttt{bing}
from Hachimori's simplicial complex library \cite{Hachimori_url}.
It is precisely on these three examples that the greedy heuristics of
Lewiner et al.\ produce somewhat inconsistent results.

In \cite{LewinerLopesTavares2003a}, $(1,1,1,0)$ was obtained for
\texttt{bing} and \texttt{knot}. In~\cite{Lewiner2005} on p.~92,
\texttt{bing} and \texttt{knot} appear with $(1,0,0,0)$
without mentioning the improvement with respect to \cite{LewinerLopesTavares2003a}.
Moreover, \texttt{nc\_sphere} is listed on p.~92 of~\cite{Lewiner2005} with $(1,2,2,1)$ 
and it is noted on p.~89:
``Trickier, the non-shellable $3$-sphere (NC Sphere) is a delicate model 
since no discrete Morse function can reach the minimal number 
of critical points for smooth homotopy.'' This latter statement is false
as we found (in 12 out of 10000 runs) the optimal discrete Morse
vector $(1,0,0,1)$ for \texttt{nc\_sphere}; cf.\ Table~\ref{tbl:discrete_morse_spectra}.
In fact, the $3$-sphere \texttt{nc\_sphere} with $f=(381,2309,3856,1928)$ is obtained 
from the $3$-ball \texttt{knot} with $f=(380,1929,2722,1172)$ by adding a cone
over the boundary of \texttt{knot}. By this, every discrete Morse function
on \texttt{knot} with discrete Morse vector $(1,0,0,0)$ can be used to
produce a discrete Morse function with discrete Morse vector $(1,0,0,1)$
on \texttt{nc\_sphere}. In contrast, it would theoretically be possible
to have \texttt{knot} with optimal discrete Morse vector $(1,1,1,0)$,
while \texttt{nc\_sphere} has optimal discrete Morse vector $(1,0,0,1)$.
The best discrete Morse vector we found in 1000000 runs for \texttt{knot} is $(1,1,1,0)$; 
see Table~\ref{tbl:discrete_morse_spectra}  ---  whereas, as mentioned above, Lewiner~\cite{Lewiner2005}
seemed to claim $(1,0,0,0)$ for this example, which would beat our algorithm.

\enlargethispage*{4mm}

\section{Theoretical lower bounds for discrete Morse vectors}
\label{sec:lower bounds}

In this section, we briefly recall some theoretical lower bounds for minimal discrete Morse vectors. The obstructions for the existence of discrete Morse functions with a certain number of critical cells are of various nature. Here we basically use four different criteria. 
The first concerns ridge-facet incidences, the second follows from elementary algebraic topology (applied to the Morse complex), 
the third uses knot theory, and the fourth comes from smooth Morse theory. 

\subsection{Ridge-facet incidences and Euler characteristic}

In order for a collapse to start, there need to be free faces. 
This is how to create a first obstruction, namely by constructing $d$-dimensional triangulations
in which every $(d-1)$-face is contained in two or more $d$-faces. 

The most famous example
of this type is the dunce hat, a contractible $2$-complex obtained from a single triangle by identifying all three boundary edges
in a non-coherent way. In any triangulation of the dunce hat each edge belongs to 
either two or three triangles; cf.~\cite{BenedettiLutz2009pre}.
Hence, the dunce hat cannot be collapsible or, in other words, it cannot have $(1,0,0)$ as discrete Morse vector.

The vectors $(1,0,1)$ and $(1,1,0)$ are also forbidden for the dunce hat.
In fact, since each elementary collapse deletes two faces of consecutive dimension,
it does not change the Euler characteristic. In particular, the alternating sum
of the entries of a discrete Morse vector should always be equal to the 
Euler characteristic of a complex. 

The dunce hat does, however, admit $(1,1,1)$
as  discrete Morse vector, which is therefore optimal.

\subsection{The Morse complex}

Forman showed that any discrete Morse vector on a simplicial complex $C$ is also the face-vector of a \emph{model} for $C$, that is, a CW-complex homotopy equivalent to $C$.
 
 \begin{thm}[{Forman~\cite{Forman2002}}]
Assume that some $d$-complex $C$ admits a discrete Morse function with $c_i$ critical faces of dimension $i$ ($i=0,\ldots, d$). Then $C$ has a model with $c_i$ $i$-cells, called \emph{Morse complex}.
\end{thm}

This theorem results in several obstructions. First of all, the $i$-th (rational) Betti number of an arbitrary CW-complex is always bounded above by its number of $i$-dimensional cells. 

\begin{cor}[Forman's weak Morse inequalities~\cite{Forman2002}] \label{cor:weakMorse}
Assume that some $d$-complex $C$ admits a discrete Morse function with $c_i$ critical faces of dimension $i$ ($i=0,\ldots, d$). Then $c_i \ge \beta_i (C)$ for each $i$.
\end{cor}

The previous result still holds if we consider homology over a finite field. 

\begin{cor} Assume some $d$-complex $C$ admits a discrete Morse function with $c_i$ critical faces of dimension $i$ ($i=0,\ldots, d$). Then $c_i \ge \dim H_i (C; \mathbb{Z}_p)$ for each $i$ and for each prime~$p$.
\end{cor}

Sometimes it is convenient to focus on homotopy groups rather than on homology groups. 
Recall that the fundamental group of a CW-complex with one $0$-cell is completely determined by its $2$-skeleton; a presentation of the group can be obtained using the $1$-cells as generators and the $2$-cells as relators. In particular, if the CW-complex  has no $1$-cells, its fundamental group must be trivial; and if  the CW-complex  has only one $1$-cell, its fundamental group must be trivial or cyclic. 

\begin{cor} Assume some $d$-complex $C$ with fundamental group $G$ admits a discrete Morse function with $1$ critical face of dimension $0$ and $c_1$ critical faces of dimension $1$. Then $c_1 \ge \operatorname{rank}(G)$, the minimal number of generators in a presentation of $G$. (In particular, if $G$ is non-abelian, then $c_1 \ge 2$.) 
\end{cor}

\subsection{Knot-theoretic obstructions}

Obstructions coming from short knots have been considered first by Bing~\cite{Bing1964}, Goodrick~\cite{Goodrick1968}, 
and Lickorish~\cite{Lickorish1991}, and later investigated 
by the two authors~\cite{Benedetti2012,BenedettiLutz2013apre,Lutz2004b} and others. 
Recall that a knot $K$ inside a triangulation of a $3$-sphere is just a $1$-dimensional subcomplex homeomorphic to a $1$-sphere (or in other words, a closed path in the $1$-skeleton.) The \emph{knot group} is the fundamental group of the knot complement inside the sphere. Knot groups are perhaps the main invariant in knot theory. 

In the simplest form (that is, for $3$-dimensional spheres) the obstructions are of the following type:

 \begin{thm}[Lickorish~\cite{Lickorish1991}; cf.\ also \cite{Benedetti2012}]\label{thm:Lickorish}
Assume some triangulated $3$-sphere $S$ admits some discrete Morse function with $c_2$ critical $2$-faces. Then, for \emph{any} knot $K$ inside $S$, one has
\[ c_2 \; \ge \; \operatorname{rank}(G_K) - f_1(K),\]
where $G_K$ is the knot group of $K$ and $f_1(K)$ is the number of edges of $K$. 
\end{thm}

The previous theorem is usually applied together with the following two well-known facts: 
\begin{compactenum}[(1)]
\item there are knots whose groups have arbitrarily high rank; for example, the knot group of a connected sum of $m$ trefoils has rank $\ge m+1$ (Goodrick~\cite{Goodrick1968});
\item any knot can be realized with only $3$ edges in a suitably triangulated $3$-sphere (Bing~\cite{Bing1964}).
\end{compactenum}

In particular, if we consider a $3$-sphere $S$ containing the connected sum of three trefoils realized on three edges, 
then Theorem~\ref{thm:Lickorish} yields $c_2 \ge 1$ for all discrete Morse vectors $(1, c_1, c_2, 1)$. Note that $c_1= c_2$, because of Euler characteristic reasons.

A similar statement can be proven for $3$-dimensional balls.

\begin{thm}[\mbox{\cite[Corollary 4.25]{Benedetti2012}}]  
\label{thm:benedetti_4_25}
Assume some triangulated $3$-ball $B$ admits some discrete Morse function with $c_1$ critical edges. Let $K$ be a knot in the $1$-skeleton of $B$, realized as a path of $b$ edges in the boundary of $B$ plus a path of\, $e=f_1(K)-b$  interior edges. Then 
\[ c_1 \; \ge \; \operatorname{rank}(G_K) - 2e,\]
where $G_K$ is the knot group of $K$. 
\end{thm}

\subsection{Morse-theoretical obstructions} 
Very recently, the first author proved the following result for smooth manifolds.

\begin{thm} (\cite{Benedetti2012pre})
Every smooth Morse vector is also a discrete Morse vector on some (compatible) PL triangulation. 
In dimensions up to $7$, the converse holds too.
\end{thm}

The converse statement is interesting for us because it yields further obstructions. For example, we know from the work by 
Boileau and Zieschang~\cite{BoileauZieschang1984} and others, that for every $r>0$, there is a (smooth) $3$-manifold $M_r$ of Heegaard genus $g\geq \operatorname{rank} (M_r) + r$.  It follows that \emph{for every PL triangulation $T$ of $M_r$}, every discrete Morse vector on $T$ has $c_1 \ge g\geq \operatorname{rank} (M_n)+r$ critical edges.

\section{Towards a new library of triangulations}
\label{sec:computational_results}

Table~\ref{tbl:discrete_morse_spectra} provides a library of $45$ instances
for which we sampled the discrete Morse spectrum. 
We ran our random algorithm 10000 rounds on each example, except for eight 
examples for which we did fewer runs. The $45$ examples were selected 
for different reasons as we will explain below. The respective examples are
listed at the beginning of each subsection.
The library of examples can be found online at \cite{BenedettiLutz_LIBRARY}.

An additional infinite series of complicated triangulations, based on a handlebody construction 
of Akbulut and Kirby \cite{AkbulutKirby1985}, was recently given in \cite{TsurugaLutz2013ext}.

\subsection{`Trivial' triangulations}
\label{sec:trivial}

\texttt{Examples:}  \texttt{dunce\_hat,} \texttt{d2n12g6,}  \texttt{regular\_2\_21\_23\_1} 

\medskip

\noindent
Discrete Morse theory is trivial on $1$-dimensional complexes (graphs) 
and $2$-dimensional compact manifolds (surfaces); cf.~\cite{LewinerLopesTavares2003a}.
A simple modification of our heuristic allows to incorporate this as follows.
Once we reduced a simplicial complex to a $1$-dimensional complex,
we switch to a deterministic strategy:  As~long as there are edges 
that are contained in a cycle, delete such (critical) edges iteratively;
then collapse the remaining tree/forest to a point/a collection of points, respectively. 

\begin{deff}
  Let\, $C$ be a connected finite simplicial complex.
  The  \emph{normalized discrete Morse spectrum} $\sigma_N$ of $C$ is obtained from the discrete Morse spectrum $\sigma$ of\, $C$
  by normalizing every discrete Morse vector $(c_0, c_1, c_2, \ldots, c_d)$ in the spectrum
   to $(1, c_1-c_0+1, c_2, \ldots, c_d)$ and adding up the probabilities for the original vectors
   that have the same normalization.
 \end{deff}
 
\emph{Example}: The graph $A_7$ of Figure~\ref{fig:A7} has normalized discrete Morse spectrum $\{\mbox{$1$-$(1,2)$}\}$
                                or, for short, $\{(1,2)\}$.

\bigskip

We introduce the following averages:
\begin{compactitem}
\item $c_{\sigma}$, the average number of critical cells for the vectors in the discrete Morse spectrum $\sigma$ of a simplicial complex $C$;
\item $c^N_{\sigma}$, the average number of critical cells for the vectors in the normalized discrete Morse spectrum $\sigma_N$ of  $C$.
\end{compactitem}
By Corollary~\ref{cor:weakMorse} we have $$c_{\sigma}\,\geq\, c^N_{\sigma}\,\geq\, \beta_0+\beta_1+\dots+\beta_d.$$

The coefficient $c^N_{\sigma}$ (and also $c_{\sigma}$)  is of some interest if we want to randomly reduce the size of a complex 
as a preprocessing step for homology computations; it gives  an estimate for the number of cells that we are left
with for the Smith normal form computations. 
                                
\begin{lem}
Every connected simplicial $1$-complex $K$ with $n$ vertices and $m\geq n-1$ edges has normalized discrete Morse spectrum
$\{(1,m-n+1)\}$ and\, $c^N_{\sigma}=2+m-n$.
\end{lem}

The homology vector  in this case is $H_*(K)=({\mathbb Z},{\mathbb Z}^{m-n+1})$,
so the weak discrete Morse inequalities (see Section~\ref{sec:classical}) are sharp.

\begin{lem}
Every triangulation $K$ of a closed (connected) surface of Euler characteristics $\chi$ has normalized discrete Morse spectrum
$\{(1,2-\chi,1)\}$ and\, $c^N_{\sigma}=4-\chi$. More generally, the same holds for every strongly connected $2$-complex $K$ 
without free edges.
\end{lem}

\emph{Proof}: Triangulations of surfaces are strongly connected. Hence, after the removal of only one critical triangle 
the remaining complex collapses onto a $1$-dimensional complex, i.e., $c_2=1$. 
The conclusion follows from the previous lemma and extends to all strongly connected $2$-complex $K$ without free edges.\hfill $\Box$

\bigskip

\noindent
The example \texttt{d2n12g6} in Table~\ref{tbl:discrete_morse_spectra}  is the unique vertex-transitive, 
vertex-minimal neighborly triangulation of  the orientable surface of genus $6$ \cite{AltshulerBokowskiSchuchert1996},
the example \texttt{regular\_2\_21\_23\_1} is a regular triangulation of the orientable surface of genus $15$
with $21$ vertices \cite[Ch.~5]{Lutz1999}.

Something can also be said on complexes with few vertices:

\begin{thm} \label{thm:bagchi_datta} 
{\rm (Bagchi and Datta~\cite{BagchiDatta2005})}
Every ${\mathbb Z}_2$-acyclic simplicial complex with at most $7$~vertices is collapsible.
\end{thm}
\begin{cor} 
Every ${\mathbb Z}_2$-acyclic simplicial complex $K$ with at most $7$ vertices is extendably collapsible
and therefore has trivial discrete Morse spectrum $\{(1,0,0)\}$ with $c_{\sigma}=c^N_{\sigma}=1$.
\end{cor}

The $7$-vertex bound is sharp; the triangulation \texttt{dunce\_hat} (cf.~\cite{BenedettiLutz2009pre})
of the dunce hat is an $8$-vertex example of a non-collapsible contractible complex.

\subsection{3-manifolds with up to ten vertices}
\label{sec:small_3manif}

\begin{table}
\small\centering
\defaultaddspace=0.15em
\caption{Total time to find optimal discrete Morse functions (in a single run) for each of the combinatorial $3$-manifolds with up to $10$ vertices.}\label{tbl:ten3d_leq10}
\begin{tabular*}{\linewidth}{@{\extracolsep{\fill}}c@{\hspace{5mm}}r@{\hspace{5mm}}r@{\hspace{5mm}}r@{\hspace{5mm}}r@{\hspace{5mm}}r@{}}
\\\toprule
 \addlinespace
 \addlinespace
  Vertices$\backslash$Types & $S^3$ & $S^2\hbox{$\times\hspace{-1.62ex}\_\hspace{-.4ex}\_\hspace{.7ex}$}S^1$ & $S^2\!\times\!S^1$ & All & Total time \\ 
                            &       &           &                &  &(in Min:Sec.Frac)      \\ \midrule
 \addlinespace
 \addlinespace
      5 &           1 &      --  &       -- &           1 & 0.008 \\[-1mm]
 \addlinespace
      6 &           2 &      --  &       -- &           2 & 0.008 \\[-1mm]
 \addlinespace
      7 &           5 &      --  &       -- &           5 & 0.012 \\[-1mm]
 \addlinespace
      8 &          39 &      --  &       -- &          39 & 0.060 \\[-1mm]
 \addlinespace
      9 &       1\,296 &        1 &       -- &       1\,297 & 3.836 \\[-1mm]
 \addlinespace
     10 &     247\,882 &      615 &      518 &     249\,015 & 17:35.606 \\
\bottomrule
\end{tabular*}
\end{table}

For all $250\,359$ examples in the catalog \cite{Lutz2008a} of triangulations of $3$-manifolds with up to $10$~vertices,
optimal discrete Morse vectors were found by a \emph{single run} of our program each;
see Table~\ref{tbl:ten3d_leq10}. 

\begin{thm}
All\, $250\,359$ examples of triangulated $3$-manifolds with up to $10$~vertices 
admit a perfect discrete Morse function.
\end{thm}

The spheres in this list are all shellable, as are all $3$-spheres 
with up to $11$ vertices~\cite{SulankeLutz2009}. The smallest known non-shellable $3$-sphere  \texttt{S\_3\_13\_56} (\texttt{trefoil})
has $13$ vertices~\cite{Lutz2004b}. For all the 1134 non-spherical examples the statement of the theorem is new.

\enlargethispage*{5mm}

\subsection{Polytopal spheres}

\texttt{Examples:}  \texttt{S\_3\_100\_4850,} \texttt{600\_cell,} \texttt{S\_3\_1000\_2990,} \texttt{S\_5\_100\_472.}

\medskip

\noindent
We ran our program 
on the $3$-dimensional boundary \texttt{S\_3\_100\_4850} of the cyclic $4$-polytope with $100$ vertices and $4850$ facets, 
on the $3$-dimensional boundary \texttt{600\_cell} of the $600$-cell,
on the $3$-dimensional boundary \texttt{S\_3\_1000\_2990} of a stacked $4$-polytope with $1000$ vertices and $2990$ facets,
and on the $4$-dimensional boundary \texttt{S\_5\_100\_472} of a stacked $5$-polytope with $100$ vertices and $472$ facets.
In all these cases we obtained the optimal discrete Morse vector $(1,0,\dots,0,1)$  in 10000 out of 10000 tries.
We also tested various other examples 
of simplicial polytopal spheres and we always observed a trivial spectrum in these experiments. 
However, the normalized discrete Morse spectrum of simplicial polytopal spheres is not trivial in general.

\begin{thm} \label{thm:KCrowley_etal} 
{\rm (Crowley, Ebin, Kahn, Reyfman, White, and Xue~\cite{KCrowleyEbinKahnReyfmanWhiteXue2003pre})}
The $7$-simplex $\Delta_7$ with $8$~vertices contains in its $2$-skeleton an $8$-vertex triangulation of the dunce hat 
onto which it collapses.
\end{thm}

\enlargethispage*{5mm}

As a direct consequence of Theorem~\ref{thm:KCrowley_etal}, the $7$-simplex $\Delta_7$ 
is \emph{not} extendably collapsible. Therefore the spectrum of its boundary is non-trivial.
Similarly, the simplicial polytopal Gr\"unbaum--Sreed\-ha\-ran $3$-sphere No.32 on $8$ vertices,
which contains a dunce hat \cite{BenedettiLutz2009pre}, has non-trivial Morse spectrum; see 
the example \texttt{dunce\_hat\_in\_3\_ball} below.

\subsection{Random spheres}

\texttt{Example:}  \texttt{S\_3\_50\_1033.} 

\medskip

\noindent
While random surfaces can easily be generated, we lack good random models 
for $3$- or higher-dimensional manifolds; cf.~\cite{DunfieldThurston2006}.
One possible approach is to consider all triangulations of $3$-spheres 
or $3$-manifolds with a fixed number $n$ of vertices, with the uniform distribution.
While this setting is very promising for performing random experiments,
we need to get a hold on the set of all the triangulations with $n$ vertices
in the first place, a task that so far has only been solved for $3$-manifolds
with up to $11$ vertices \cite{SulankeLutz2009}.

Another model can be derived by performing random walks on the set of all triangulations where each step
is represented by a single bistellar flip.
According to a theorem of Pachner~\cite{Pachner1987}, two distinct triangulations
of a manifold are PL homeomorphic if and only if they can be connected
by a sequence of bistellar flips. 
 An implementation of bistellar flips for exploring the space of triangulations within one PL component
 is the program BISTELLAR~\cite{Lutz_BISTELLAR}; see  \cite{BjoernerLutz2000} 
for a program description. The bistellar flip approach for generating random triangulations
depends on the number of executed flips as well as on the way the flips are chosen.
As a consequence, triangulations with $n$ vertices are not selected according to the uniform
distribution. 

For the example \texttt{S\_3\_50\_1033}, we started with the boundary 
of the cyclic $4$-polytope with $50$ vertices and face vector $f=(50,1225,2350,1175)$.
We then applied $1500$ bistellar $1$-flips and reverse-$1$-flips that were chosen 
randomly from all admissible flips. The resulting sphere  \texttt{S\_3\_50\_1033} 
has $f$-vector $(50,1083,2066,1033)$. The average number of critical cells  in 10000 runs
turned out experimentally to be roughly~$3.2$ (which is considerably larger than $2$).
We therefore can conclude heuristically that random spheres tend to have a non-trivial spectrum.

\subsection{Knotted triangulations of balls and spheres}\label{sec:complicated_balls_and_spheres}

\texttt{Examples:}  \texttt{dunce\_hat\_in\_3\_ball},    
\texttt{Barnette\_sphere},
\texttt{B\_3\_9\_18},
\texttt{trefoil\_arc},
\texttt{trefoil}, \linebreak
\texttt{rudin},
\texttt{double\_trefoil\_arc},
\texttt{double\_trefoil},
\texttt{triple\_trefoil\_arc},
\texttt{triple\_trefoil}, \linebreak
\texttt{non\_4\_2\_colorable},
\texttt{knot},
\texttt{nc\_sphere},
\texttt{bing}.

\medskip

\noindent
The example \texttt{dunce\_hat\_in\_3\_ball} \cite{BenedettiLutz2009pre} is a triangulated $3$-ball that 
contains the $8$-vertex triangulation \texttt{dunce\_hat} 
in its $2$-skeleton. To indeed get stuck with \texttt{dunce\_hat},
we need to perform collapses without removing any 
of the $17$ triangles of the dunce hat.
This results in a low probability to get stuck. 
Indeed, in 1\,000\,000 runs we always found $(1,0,0,0)$
as resulting discrete Morse vector.

The non-polytopal \texttt{Barnette\_sphere} \cite{Barnette1973c} with
$8$ vertices also has trivial observed spectrum: In 1\,000\,000 runs of our program 
we obtained the optimal discrete Morse vector $(1,0,0,1)$.
For the non-shellable $3$-ball \texttt{B\_3\_9\_18} \cite{Lutz2004a}
with $9$ vertices and Rudin's non-shellable $3$-ball \texttt{rudin} \cite{Rudin1958,Wotzlaw2005}
with $14$ vertices we achieved the optimal discrete Morse vector $(1,0,0,0)$ in every run.
Therefore, non-polytopality and non-shellability not necessarily cause a non-trivial observed spectrum.  

If we wish to construct triangulated balls or spheres of small size with a very non-trivial observed spectrum,
we need to build in complicated substructures 
of small size (like complicated knots on few edges) to get stuck at.

The triangulated $3$-sphere \texttt{trefoil}  (\texttt{S\_3\_13\_56} \cite{Lutz2004b})
contains a $3$-edge trefoil knot in its $1$-skeleton 
and has optimal discrete Morse vector $(1,0,0,1)$. 
This vector was obtained in roughly 96\% of the runs of our heuristic.
The triangulated $3$-sphere \texttt{double\_trefoil} (\texttt{S\_3\_16\_92} \cite{BenedettiLutz2013apre})
with optimal discrete Morse vector $(1,0,0,1)$
has a $3$-edge double trefoil knot in its $1$-skeleton. Here, $(1,0,0,1)$ was achieved only in 40\%
of the runs.
The triangulated $3$-sphere \texttt{triple\_trefoil} (\texttt{S\_3\_18\_125} \cite{BenedettiLutz2013apre})
contains a $3$-edge triple trefoil knot in its $1$-skeleton 
and has optimal discrete Morse vector $(1,1,1,1)$, which we found 30\% of the time.

The $3$-ball \texttt{trefoil\_arc} is obtained
from the $3$-sphere \texttt{trefoil} by deleting the 
star of a vertex. It contains the trefoil knot as a spanning arc
and has optimal discrete Morse vector $(1,0,0,0)$.
The deletion of the star of a vertex from the $3$-sphere \texttt{double\_trefoil}
yields the $3$-ball  \texttt{double\_trefoil\_arc}
with the double trefoil knot as spanning arc and optimal discrete Morse vector $(1,1,1,0)$.

For the triple trefoil knot the deletion of a vertex from the $3$-sphere 
\texttt{triple\_trefoil} yields the $3$-ball \texttt{triple\_trefoil\_arc} 
for which the optimal discrete Morse vector is $(1,2,2,0)$; see Theorem~\ref{thm:benedetti_4_25}.
We found this vector in about 60\% of the runs.

A larger $3$-ball \texttt{knot} that has the trefoil knot as spanning arc
was constructed (via a pile of cubes) by Hachimori \cite{Hachimori_url}.
The best discrete Morse vector we found for \texttt{knot} is $(1,1,1,0)$.
It might as well be that \texttt{knot} admits $(1,0,0,0)$ as optimal discrete Morse vector.
The non-constructible $3$-sphere  \texttt{nc\_sphere} \cite{Hachimori_url}
is obtained from \texttt{knot} by adding the cone over the boundary of \texttt{knot}.
For this example, we found $(1,0,0,1)$ as optimal discrete Morse vector,
but only in 12 out of 10000 runs. 

The triangulation \texttt{bing} is a $3$-dimensional thickening of Bing's house with two rooms \cite{Bing1964}
due to Hachimori \cite{Hachimori_url} (again, via a pile of cubes). It is a $3$-ball with $480$ vertices 
for which we found $(1,0,0,0)$ as optimal discrete Morse vector in
only $7$ out of 10000 runs. We therefore can regard this ball as
\emph{barely collapsible}.

A non-$(4,2)$-colorable triangulation \texttt{non\_4\_2\_colorable} of the $3$-sphere
was constructed in \cite{LutzMoller2013pre}
with 167 vertices by using 10 copies of the double trefoil knot.
The best discrete Morse vector we found once for this example in 10000 runs
is $(1,2,2,1)$. The average number of critical cells for  \texttt{non\_4\_2\_colorable}  computed and normalized 
over only 10 random runs (for the sake of simplicity) as listed in Table~\ref{tbl:discrete_morse_spectra} is roughly 25.2.

\subsection{Barycentric subdivisions}

\begin{table}
\small\centering
\defaultaddspace=0.15em
\caption{Average number of critical cells for the knotted spheres and their barycentric subdivisions, based on 10000 random runs.}\label{tbl:barycentric}
\begin{tabular*}{\linewidth}{@{\extracolsep{\fill}}l@{\hspace{1mm}}r@{\hspace{10mm}}l@{\hspace{1mm}}r@{\hspace{10mm}}l@{\hspace{1mm}}r@{}}
\\\toprule
 \addlinespace
 \addlinespace
    \texttt{trefoil}      & $2.0778$ & \texttt{double\_trefoil}       & $3.5338$ & \texttt{triple\_trefoil}      & $5.9898$ \\[-1mm]
 \addlinespace
    \texttt{trefoil\_bsd} & $2.0202$ & \texttt{double\_trefoil\_bsd}   & $3.3414$ & \texttt{triple\_trefoil\_bsd} & $5.7352$ \\[-1mm]
 \addlinespace

\bottomrule
\end{tabular*}
\end{table}

\texttt{Examples:}   \texttt{trefoil\_bsd,}  \texttt{double\_trefoil\_bsd,}  \texttt{triple\_trefoil\_bsd.}

\medskip

\noindent
Interestingly, the barycentric subdivisions \texttt{trefoil\_bsd}, 
\texttt{double\_trefoil\_bsd}, and \linebreak
\texttt{triple\_trefoil\_bsd}
of the knotted spheres \texttt{trefoil}, \texttt{double\_trefoil}, and
\texttt{triple\_trefoil}, respectively, have a lower observed spectrum
than the corresponding original spheres; compare Table~\ref{tbl:barycentric}.

\subsection{Standard and exotic PL structures on 4-manifolds}

\texttt{Examples:}   \texttt{CP2,}  \texttt{RP4,} \texttt{K3\_16,} \texttt{K3\_17,}
                     \texttt{RP4\_K3\_17,} \texttt{RP4\_11S2xS2.}

\medskip

\noindent
Freedman's classification \cite{Freedman1982} of simply connected closed topological 
$4$-manifolds settled the $4$-dimensional topological Poincar\'e conjecture. 
The $4$-dimensional smooth Poincar\'e conjecture, however, is still wide open: 
Does the $4$-dimensional sphere $S^4$ have a unique differentiable structure 
or are there exotic $4$-spheres that are homeomorphic but not diffeomorphic to~$S^4$? 
The categories PL and DIFF coincide in dimension $4$ (see the survey 
of Milnor \cite{Milnor2011} and the references contained therein), and the $4$-dimensional smooth
Poincar\'e conjecture therefore can be rephrased on the level of triangulations:
Is every triangulation of $S^4$ PL homeomorphic to the boundary of the $5$-simplex?

Exotic structures on simply connected $4$-manifolds have been intensively studied over the past years. 
One main task has been to find smaller and smaller $k$ and $l$ such that the connected sum 
$(\# k\,{\mathbb C}{\bf P}^2)\#(-\# l\,{\mathbb C}{\bf P}^2)$ has exotic structures. 
While it is now known that ${\mathbb C}{\bf P}^2\#(-\#2\,{\mathbb C}{\bf P}^2)$~\cite{AkhmedovBDPark2010}
admits (infinitely many) exotic structures,
the remaining interesting open cases are ${\mathbb C}{\bf P}^2\#(-{\mathbb C}{\bf P})$, ${\mathbb C}{\bf P}^2$,
and $S^4$ (the smooth Poincar\'e conjecture). 

The example \texttt{CP2} in Table~\ref{tbl:discrete_morse_spectra} 
is the unique vertex-minimal $9$-vertex triangulation of ${\mathbb C}{\bf P}^2$ 
due to K\"uhnel and Banchoff \cite{KuehnelBanchoff1983}
and it carries the standard PL structure.

The constructions of exotic structures are often delicate and it is not straightforward
to derive corresponding triangulations. A very explicit example,
though not simply connected, is due to Kreck \cite{Kreck1984b}.

\begin{thm}  \label{thm:kreck} 
{\rm (Kreck~\cite{Kreck1984b})}
The $4$-dimensional manifolds\, ${\mathbb R}{\bf P}^4\#K3$\, and\, ${\mathbb R}{\bf P}^4\#(S^2\times S^2)^{\#11}$\,
are homeomorphic but not diffeomorphic; the constituting components being equipped with the standard smooth structures.
\end{thm}

A $17$-vertex triangulation  \texttt{K3\_17}  of the K3 surface with the standard PL type
is due to Spreer and K\"uhnel \cite{SpreerKuehnel2011}.  
A vertex-minimal $16$-vertex triangulation  \texttt{K3\_16} of the topological K3 surface 
was previously found by Casella and K\"uhnel \cite{CasellaKuehnel2001}.
It is not clear whether the two triangulations are PL homeomorphic ---
we tried bistellar flips to establish a PL homeomorphism between these two triangulations, 
but without success.

A vertex-minimal $16$-vertex triangulation \texttt{RP4}
of ${\mathbb R}{\bf P}^4$ was obtained in  \cite{Lutz2005bpre}  by applying bistellar flips 
to the $31$-vertex standard triangulation of ${\mathbb R}{\bf P}^4$ by K\"uhnel \cite{Kuehnel1987}.

\enlargethispage*{3mm}

Let $K$ and $L$ be two triangulated $4$-manifolds $K$ and $L$ with $n$ and $m$ vertices,
respectively. Their connected sum $K\#L$ (or $K\#-L$ in cases when
orientation of the components matters) is obtained from $K$ and $L$ by removing a $4$-simplex 
from each of the triangulations and then gluing together the remainders along the respective boundaries.
The resulting triangulation  $K\#L$ then has $n+m-5$ vertices.
Triangulations of connected sums $(S^2\times S^2)^{\# k}$, $k\geq 2$,
are therefore easily constructed from
a vertex-minimal $11$-vertex triangulation of $S^2\times S^2$ \cite{Lutz2005bpre} by taking connected sums
and then applying bistellar flips to reduce the numbers of vertices.
This way, we obtained triangulations of $(S^2\times S^2)^{\# 2}$ with $12$~vertices (vertex-minimal; c.f.\ \cite{Lutz2005bpre}),\linebreak
of $(S^2\times S^2)^{\# 3}$ with  $14$~vertices, of $(S^2\times S^2)^{\# 5}$ with $16$~vertices, 
of $(S^2\times S^2)^{\# 6}$ with  $16$~vertices, of $(S^2\times S^2)^{\# 9}$ with $18$~vertices,
and of $(S^2\times S^2)^{\# 11}$ with $20$~vertices.

\begin{thm}  \label{thm:non_PL_homeomorphic} 
Let ${\mathbb R}{\bf P}^4$, $K3$, and $(S^2\times S^2)^{\#11}$
be equipped with their standard PL structures.
The PL $4$-manifold\, ${\mathbb R}{\bf P}^4\#K3$\, has a triangulation
\texttt{RP4\_K3\_17} with $16+17-5=28$ vertices
and the PL $4$-manifold\, ${\mathbb R}{\bf P}^4\#(S^2\times S^2)^{\#11}$\, 
has a triangulation \texttt{RP4\_11S2xS2} with $16+20-5=31$ vertices. 
While the underlying topological manifolds of these PL manifolds are homeomorphic,
the respective triangulations are not PL homeomorphic.
\end{thm}

By Theorem~\ref{thm:non_PL_homeomorphic} we see that homeomorphic but not PL homeomorphic triangulations 
of $4$-manifolds can be constructed with only few vertices. (Most likely,
the explicit numbers of vertices  in Theorem~\ref{thm:non_PL_homeomorphic} 
can be further reduced with bistellar flips. However, this would require a rather extensive search, which is beyond the scope 
of this article.)

\begin{thm} 
The examples  \texttt{CP2},  \texttt{RP4}, \texttt{K3\_16},
\texttt{K3\_17}, \texttt{RP4\_K3\_17}, and \texttt{RP4\_11S2xS2}
have perfect discrete Morse functions with $3$, $5$, $24$, $24$, $27$, and $27$ critical cells, respectively.
\end{thm}

Interestingly, the computed discrete Morse spectra  of \texttt{K3\_16}  and \texttt{K3\_17} look rather similar.
The same can be said for the pair \texttt{RP4\_K3\_17}  and \texttt{RP4\_11S2xS2}.

\subsection{Hom complexes}\label{sec:hom_complexes}

\texttt{Examples:}  \texttt{Hom\_C5\_K4,} \texttt{Hom\_n9\_655\_compl\_K4,} 
                    \texttt{Hom\_C6\_compl\_K5\_small,} \texttt{Hom\_C6\_compl\_K5,} 
                    \texttt{Hom\_C5\_K5.}

\medskip

\noindent 
Hom complexes of certain graphs provide interesting examples of
prodsimplicial manifolds~\cite{CsorbaLutz2006}. The prodsimplicial
structure allows to easily triangulate these manifolds without adding 
new vertices.

The $3$-dimensional Hom complex \texttt{Hom\_C5\_K4} is a triangulation of the 
$3$-dimensional real projective space ${\mathbb R}{\bf P}^3$,
the Hom complex \texttt{Hom\_n9\_655\_compl\_K4} triangulates $(S^2\!\times\!S^1)^{\# 13}$.

The $4$-dimensional example \texttt{Hom\_C6\_compl\_K5\_small} 
with $f=(33,379,1786,2300,920)$ is obtained from \texttt{Hom\_C6\_compl\_K5} 
with $f=(1920,30780,104520,126000,50400)$ via bistellar flips.
Both examples triangulate $(S^2\!\times\!S^2)^{\# 29}$,
the first with computed normalized average  $63.92$, the latter with normalized average $83.0$.
In only three out of 2000 runs we found the discrete Morse vector $(1,1,59,0,1)$ for  \texttt{Hom\_C6\_compl\_K5},
but never the optimum $(1,0,58,0,1)$. In contrast, both the \texttt{lex}
and the \texttt{rev\_lex} heuristics yielded $(1,0,58,0,1)$.
In order to keep the list short, Table~\ref{tbl:discrete_morse_spectra} only lists 10 random runs for \texttt{Hom\_C6\_compl\_K5}.

The Hom complex \texttt{Hom\_C5\_K5} with $f=(1020,25770,143900,307950,283200,94400)$
is a triangulation of $S^3\!\times\!S^2$
with normalized average  $4.6$.

\subsection{Higher-dimensional manifolds}
\label{sec:classical}

\texttt{Examples:}  \texttt{poincare,} \texttt{hyperbolic\_dodecahedral\_space,} \texttt{S2xpoincare,}
                                \texttt{SU2\_SO3,}  \texttt{RP5\_24,}
                                \texttt{non\_PL,}  \texttt{\_HP2.}

\medskip

\noindent
The $16$-vertex triangulation \texttt{poincare} \cite{BjoernerLutz2000,BjoernerLutz2003} 
of the Poincar\'e homology $3$-sphere with $f$-vector $f=(16,106,180,90)$ 
has the binary icosahedral group as its fundamental group. Since this
group is non-cyclic, we have $c_2\geq 2$, and therefore 
every discrete Morse vector for \texttt{poincare} must have
at least six critical cells, with $(1,2,2,1)$ 
being the optimal discrete Morse vector according to
Table~\ref{tbl:discrete_morse_spectra}; cf.\ also Lewiner~\cite{LewinerLopesTavares2003b}.

For the $21$-vertex triangulation \texttt{hyperbolic\_dodecahedral\_space} \cite{LutzSulankeSwartz2009} 
of the Weber--Seifert hyperbolic dodecahedral space \cite{WeberSeifert1933} with face vector $f=(21,193,344,172)$ 
the best discrete Morse vector we found is $(1,4,4,1)$.  The fundamental group of this manifold 
can be presented with $4$ generators; see Table~\ref{tbl:gen_fund_groups}.

The product triangulation \texttt{S2xpoincare} of $S^2$ (taken as the boundary of a tetrahedron) 
with \texttt{poincare} again has the binary icosahedral group as its fundamental group, 
inherited from \texttt{poincare}; for constructing product triangulations
see \cite{Lutz2003bpre} and references therein.
The best discrete Morse vector we found for this examples in 103 out of 1000 runs is $(1,2,3,3,2,1)$.
(Table~\ref{tbl:discrete_morse_spectra} list only 20 random runs for \texttt{S2xpoincare}
to keep the list short.)

The two $5$-manifolds $SU2/SO3$ and ${\mathbb R}{\bf P}^5$
have homology vectors $(\mathbb{Z},0,\mathbb{Z},\mathbb{Z},0,\mathbb{Z})$ and 
$(\mathbb{Z},\mathbb{Z}_2,0,\mathbb{Z}_2,0,\mathbb{Z})$
and triangulations \texttt{SU2\_SO3} with $13$ vertices
and \texttt{RP5\_24} with $24$ vertices, respectively \cite{Lutz1999,Lutz2005bpre}.
The $15$-vertex triangulation \texttt{\_HP2} of an $8$-dimensional manifold 
`like a quaternionic projective plane' by Brehm and K\"uhnel \cite{BrehmKuehnel1992}
has homology $(\mathbb{Z},0,0,0,\mathbb{Z},0,0,0,\mathbb{Z})$.

\begin{thm}
The triangulations \texttt{SU2\_SO3}, \texttt{RP5\_24}, and \texttt{\_HP2}
have optimal discrete Morse vectors $(1,0,1,1,0,1)$, $(1,1,1,1,1,1)$,
and $(1,0,0,0,1,0,0,0,1)$, respectively.
\end{thm}

The $18$-vertex non-PL triangulation \texttt{non\_PL} \cite{BjoernerLutz2000} of the
$5$-dimensional sphere $S^5$ admits $(1,0,0,2,2,1)$ as discrete Morse vector.

\subsection{Random 2-complexes and fundamental groups} \label{sec:fundamental_groups}

\texttt{Example:}   \texttt{rand2\_n25\_p0.328}  

\medskip

\noindent
In generalization of the classical Erd\H{o}s--R\'enyi model for random graphs \cite{ErdosRenyi1960}, 
Linial and Meshulam \cite{LinialMeshulam2006} considered random $2$-dimensional complexes 
with complete $1$-skeleton on $n$~vertices; every triangle with vertices from the set $\{1,\dots,n\}$
is then added with probability $p$ independently. Let $Y(n,p)$ be the set of such complexes.
For the elements of $Y(n,p)$, Linial and Meshulam proved a sharp threshold for the vanishing of the first homology
with ${\mathbb Z}_2$-coefficients,
$$\lim_{n\rightarrow\infty}\,{\rm Prob}[\,Y\in Y(n,p)\,|\,H_1(Y,{\mathbb Z}_2)=0\,]
=\left\{\begin{tabular}{lll}$1$ & {\rm for} & $p=\frac{2{\rm log}n+\omega(n)}{n}$,\\[2mm] 
                            $0$ & {\rm for} & $p=\frac{2{\rm log}n-\omega(n)}{n}$,
                                            \end{tabular}\right.$$
for any function $\omega(n)\rightarrow\infty$ as $n\rightarrow\infty$ (as long as $p\in [0,1]$).
Replacing homological connectivity by simple connectivity, Babson, Hoffman, and Kahle \cite{BabsonHoffmanKahle2010}
showed that there is a range for $p$ for which asymptotically almost surely the complexes $Y\in Y(n,p)$ have non-trivial
fundamental groups with trivial abelianizations,
$$\lim_{n\rightarrow\infty}\,{\rm Prob}[\,Y\in Y(n,p)\,|\,\pi_1(Y)=0\,]\,=1 \quad {\rm for}\quad p\geq \displaystyle\big(\textstyle\frac{3{\rm log}n+\omega(n)}{n}\displaystyle\big)^{\frac{1}{2}},$$
with the exponent $\frac{1}{2}$ being best possible.

More recently, Cohen et al.~\cite{DCohenCostaFarberKappeler2012} showed that for 
$p\ll n^{-1}$ asymptotically almost surely the complexes $Y\in Y(n,p)$ admit a discrete Morse functions
with no critical $2$-cells. See also the recent results in higher dimensions by Aronshtam, Linial, {\L}uczak and Meshulam~\cite{AronshtamLinialLuczakMeshulam2013}, where they consider the case $p=c\cdot n^{-1}$.

\enlargethispage*{3mm}

The example  \texttt{rand2\_n25\_p0.328}  on $n=25$ vertices from Table~\ref{tbl:discrete_morse_spectra} 
with homology $(\mathbb{Z},0,\mathbb{Z}^{475})$ has $751$ triangles, each picked with probability $p=0.328$. 
We found the optimal discrete Morse vector $(1,0,475)$ in $275$ out of 10000 runs.

According to Seifert and Threlfall \cite[\S 44]{SeifertThrelfall1934}, a presentation of the fundamental group
of a simplicial complex can be obtained via the edge-path group. For this, a spanning tree of edges
is deleted from the 1-skeleton of the complex and each remaining edge contributes a generator to the 
fundamental group, while each triangle of the 2-skeleton contributes a
relator; see \cite{Lutz_FundamentalGroup} for an implementation.

We used the GAP command \texttt{SimplifiedFpGroup} to simplify the edge-path group presentation 
of the fundamental group. The heuristic  \texttt{SimplifiedFpGroup}
not necessarily outputs a minimal presentation of a finitely presented
group with the minimal number of generators and relators.
Nevertheless, even in the case of huge complexes, \texttt{SimplifiedFpGroup} 
succeeded with recognizing trivial, cyclic (one generator, at most one
relator), or free groups (no relators). 
In Table~\ref{tbl:gen_fund_groups}, we list for the examples of Table~\ref{tbl:discrete_morse_spectra} 
the number of generators (Ge.)~and the number of relators (Re.)~of the initial presentation
and the number of generators (SGe.)~and the number of relators (SRe.)~of the simplified group
along with the resulting fundamental group (F.~Gr.)~and the time it took for the simplification.
In Tables~\ref{tbl:gen_fund_groups}--\ref{tbl:rand_2compl_n50}, 
$F(k)$ denotes the free group with $k$ generators.

In the Tables~\ref{tbl:rand_2compl_n25}  and~\ref{tbl:rand_2compl_n50}
we list resulting fundamental groups for random $2$-complexes with 25
and 50 vertices, respectively. In these tables, the Linial--Meshulam
threshold can be observed quite nicely. For $p=\frac{2{\rm log}25}{25}\approx 0.26$
and $p=\frac{2{\rm log}50}{50}\approx 0.16$, 73 and 75 out of 100 random
examples with $n=25$ and $n=50$ vertices had trivial fundamental groups, respectively. 
Thus, for these values of $p$ we precisely are in the range of the slope of the threshold.
While most of the examples in the Tables~\ref{tbl:rand_2compl_n25} and~\ref{tbl:rand_2compl_n50} 
have free fundamental groups, we found `non-free' examples (for which their
presentations could not be simplified to remove all relators) in the
range when $p$ is slightly smaller than $\frac{3}{n}$, the value for which
Linial, Meshulam, and Rosenthal~\cite{LinialMeshulamRosenthal2010}
constructed acyclic examples as sum complexes.

In our experiments we did not observe the Babson--Hoffmann--Kahle examples 
with non-trivial fundamental groups that have trivial abelianizations.
However, as pointed out by Kenyon, `exceptional events' can occur 
for random groups in the case when $n$ is small while
the asymptotical behavior can be rather different; cf.~\cite[pp.~42--43]{Ollivier2005}.

\subsection{Vertex-homogeneous complexes and the Evasiveness Conjecture}

\texttt{Example:} \texttt{contractible\_vertex\_homogeneous.}

\medskip

\noindent
As remarked by Kahn, Saks, and Sturtevant \cite{KahnSaksSturtevant1984} we have 
the following implications for simplicial complexes:
$$\mbox{non-evasive}\quad\Longrightarrow\quad\mbox{collapsible}\quad\Longrightarrow\quad\mbox{contractible}\quad\Longrightarrow\quad\mbox{${\mathbb Z}$-acyclic}.$$

The Evasiveness Conjecture \cite{KahnSaksSturtevant1984}  for simplicial complexes states that 
every vertex-homo\-ge\-neous non-evasive simplicial complex is a simplex.
The first examples of vertex-homo\-ge\-neous ${\mathbb Z}$-acyclic simplicial complexes different from simplices
were given by Oliver (cf.~\cite{KahnSaksSturtevant1984}); see~\cite{Lutz2002b} for further examples.
While join products and other constructions can be used to  derive vertex-homo\-ge\-neous contractible
simplicial complexes different from simplices, non-trivial  vertex-homo\-ge\-neous non-evasive examples
cannot be obtained this way~\cite{Welker1999}. 

The smallest example \texttt{contractible\_vertex\_homogeneous} of a contractible vertex-homo\-ge\-neous
simplicial complex from \cite{Lutz2002b} is $11$-dimensional with
$$f=(60,1290,12380,58935,148092,220840,211740,136155,59160,16866,2880,225).$$          
The best discrete Morse vector we found with the \texttt{lex}
and the \texttt{rev\_lex} heuristics for this contractible space is $(1,0,0,4,8,4,0,0,0,0,0,0)$.
We do not know whether the example is collapsible or not.

\subsection*{Acknowledgments}

Thanks to Karim Adiprasito, Herbert Edelsbrunner, Alex Engstr\"om, Michael Joswig, Roy Meshulam, Konstantin Mischaikow, 
Vidit Nanda and John M.~Sullivan for helpful discussions and remarks.

\section*{Appendix A: Complexes on which our heuristic fails}
In this section, we construct simplicial complexes on which our random approach 
will most likely go far from guessing the right Morse vector.
On these examples, exponentially many rounds (in the number of facets) 
of the program may be necessary, before an optimal Morse vector shows up as output. 
Such pathological examples can be produced in any positive dimension. 
The crucial idea is highlighted by the following $1$-dimensional case.

\begin{figure}
\begin{center}
\includegraphics[width=6.5cm]{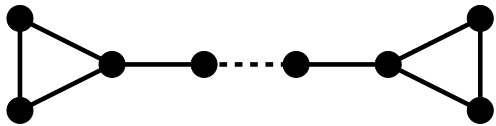}
\end{center}
\caption{The graph $A_{k+6}$ with $k+6$ vertices.}
\label{fig:Ak6}
\end{figure}

\begin{example} \label{ex:Bad1} \rm
Let $k$ be a positive integer. Let $A_{k+6}$ be the graph consisting of two cycles 
of length $3$ that are connected by a path of $k$ edges; see Figure~\ref{fig:Ak6}.
Since $A_{k+6}$ has no free edge, our algorithm picks an edge $e$ uniformly at random and
removes it. The final outcome depends on this choice, and on this choice only:
\begin{compactitem}
\item If $e$ belongs to the $k$-edge path, it is easy to see that the program will always output the discrete Morse vector~$(2,3)$. 
\item If instead $e$ belongs to one of the two triangles, then the program 
  will always output the Morse vector $(1,2)$. 
\end{compactitem}
Hence, the algorithm finds a perfect
Morse function on $A_{k+6}$ with probability $p=\frac{6}{6+k}$.
For large $k$,  the algorithm will most likely (i.e. with
probability $q=\frac{k}{6+k}$) return a Morse vector that is `off by $2$', displaying $5$ critical cells instead of $3$. 
\end{example}

\enlargethispage*{12mm}

\begin{example} \label{ex:Bad2} \rm
Let $s$ be a positive integer. Let $B_{k+6} (s)$ be a bouquet of $s$ copies of
$A_{k+6}$. An optimal discrete Morse function on $B_{k+6} (s)$ has Morse vector $(1,2s)$. 
Finding a discrete Morse function on $B_{k+6} (s)$ is the same as (independently) finding 
$s$ discrete Morse functions on the $s$ copies of $A_{k+6}$. Therefore, the probability 
of getting the optimal Morse vector on $B_{k+6} (s)$ is $p^s$, where $p=\frac{6}{6+k}$. 
This corresponds to putting together $s$ optimal Morse functions on the different copies 
of $A_{k+6}$, or in other words, to picking one favorable edge in each copy of $A_{k+6}$.
For $0 \le i \le s$, the probability that the program outputs the Morse vector $(1+i,2s+i)$\, 
is\, $\binom{s}{i} p^{s-i} (1-p)^i$, corresponding to $i$ `bad choices' and $s-i$ `good choices'.
\end{example}

To show that an analogous phenomenon occurs also in higher dimensions, let us recall a classical definition in PL topology.

\begin{deff} \rm
Let $C$ be a $d$-dimensional complex. A \emph{stacking operation}
on $C$ is the transition from $C$ to
$ C' =  (C - \operatorname{star}(\sigma, C) \, ) \; \cup \; \hat{
\sigma } \ast \operatorname{link}(\sigma, C)$,
where $\sigma$ is an arbitrary facet of $C$ and $\hat{\sigma}$ is a new vertex 
(e.g., the barycenter of $\sigma$). More generally, we say that $C'$ {\em is obtained from $C$ by stacking} 
if some finite sequence of stacking operations leads from $C$ to $C'$.
\end{deff}

Each stacking operation adds $d$ facets; 
so, a complex obtained by performing $s$ stacking operations on a $d$-simplex has exactly $ds + 1$ facets.  

\begin{lem}
If $C'$ is obtained from a simplex by stacking, then $C'$ is shellable. In particular, 
it is endo-collapsible: For any facet $\sigma$ of $C'$, there is a sequence of elementary collapses 
that reduces $C' - \sigma$ to $\partial C'$. 
\end{lem}

In dimension $d \ge 3$, there is no guarantee that \emph{any} sequence of elementary collapses 
on $C' - \sigma$  can be continued until one reaches $\partial C'$. 
This is why, in the following example, probabilities have to be estimated rather than computed.

\begin{figure}[t]
\begin{center}
\includegraphics[width=4cm]{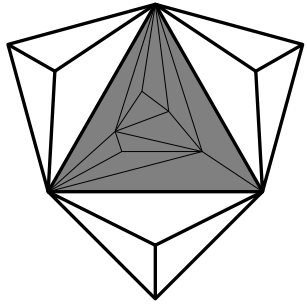}
\end{center}
\caption{An example $C^2_{2\cdot 5+1}$ with the central $2$-simplex $S$ (in grey) 
              subdivided $5$ times and the boundary edges of $S$ blocked by three empty tetrahedra.}
\label{fig:C_2_11}
\end{figure}

\begin{example} \label{ex:Bad3} \rm
Let $d, k$ be positive integers, with $k \equiv 1$ mod $d$. Take a
disjoint union of $d+1$ edges $e_0, \ldots, e_d$, and a $d$-simplex $S$
(with facets $F_0, \ldots, F_d$). For each $i$ in $\{0, \ldots, d\}$, glue
in the boundary of the join $F_i \ast e_i$. The resulting complex $C^d$ is
homotopy equivalent to a bouquet of $d+1$ spheres of dimension $d$; the
homotopy is just the contraction of the central simplex $S$ to a point. 
Let $C^d_k$ be a complex obtained by stacking the simplex $S$ exactly $s$~times, 
so that $S$ gets subdivided into $k = d s + 1$ simplices of the same dimension. 

Note first of all that $C^1_k$ coincides with the $A_{k+6}$ of Example~\ref{ex:Bad1}. 
For an example $C^2_{2\cdot 5+1}$ see Figure~\ref{fig:C_2_11}.
Since $C^d_k$ has no free $(d-1)$-faces, our algorithm starts by removing some $d$-face $\sigma$ at random. 
We have two possible cases:
\begin{compactitem}
\item With probability $\frac{k}{(d+2)(d+1) + k}$ we pick $\sigma$ from
the subdivision of the central simplex. 
\item With probability $\frac{(d+2)(d+1)}{(d+2)(d+1) + k}$ we pick
$\sigma$ from one of the $d$-spheres. 
\end{compactitem}

In the first case, \emph{some} sequence of elementary collapses reduces $C^d_k - \sigma$ 
onto $ C^d - S$. So our algorithm will output a Morse vector 
that is either $(1, 0, \ldots, 0, 1, d+2)$ or a (componentwise) larger vector; 
but certainly not the vector $(1,0, \ldots, 0, 0, d+1)$.

Thus the probability of obtaining the optimal Morse vector $(1,0, \ldots, 0, 0, d+1)$ 
is positive, but smaller or equal to $\frac{(d+2)(d+1)}{(d+2)(d+1) + k}$. 
As $k$ gets larger, this upper bound gets smaller and smaller.
\end{example}

\begin{example} \rm
By taking a bouquet of $w$ copies of Example \ref{ex:Bad3}, we obtain a
complex $B^d_k (w)$. For $d=1$, $B^d_k (w)$ coincides with  the $B_{k+6} (w)$ of  Example \ref{ex:Bad2}.
The probability of seeing the perfect Morse vector $(1, 0, \ldots, 0, 0, (d+1)w)$ on $B^d_k (w)$ 
is smaller or equal to $\left( \frac{(d+2)(d+1)}{(d+2)(d+1) + k} \right)^w.$ 
\end{example}

For practical purposes, it is useful to understand how this probability grows 
\emph{with respect to the number $N$ of facets}. In fact, given a complex with $N$ facets, 
we would like to concretely know how often we should run the algorithm before 
we can expect an optimal Morse vector to appear among the outputs. 

\enlargethispage*{1mm}

For the sake of brevity, we do
the calculations in dimension one --- but similar estimates can be easily
derived in all dimensions. The graph constructed in Example~\ref{ex:Bad2}
has $N=(6+k)w$ edges. To study the probability $(\frac{6}{6+k})^w$ of
finding an optimal Morse function, we should regard $N$ as a constant,
write $w$ as $\frac{N}{6+k}$, and study the function
\[ P(k) = \left( \frac{6}{6+k} \right) ^{\frac{N}{6+k}}.\]
Now, classical calculus reveals that the function $x \longmapsto x^x =
e^{x \log x}$ is strictly decreasing on the interval $(0,e^{-1})$ and
strictly increasing on $(e^{-1}, \infty)$. It achieves its minimum at
$e^{-1}$. So, given any bijection $g: (0, \infty) \rightarrow (0,1)$, the
function $y \longmapsto g(y)^{g(y)}$ achieves its minimum at the (unique) point $y$
such that $g(y)=e^{-1}$. Applying this to 
$g(y)=\frac{6}{6+y}$, we get
\[ \min_{y \in \mathbb{R}} \left( \frac{6}{6+y} \right) ^{\frac{N}{6+y}}
=
\min_{y \in \mathbb{R} } \left( g(y)^{g(y) \frac{N}{6}} \right)
=
\left( \min_{y \in \mathbb{R} } \, g(y)^{g(y) } \right)^{\frac{N}{6}} =
\left( \, (e^{-1})^{e^{-1}} \right)^{\frac{N}{6}} = \ e^{-\frac{N}{6e}}.\]
Yet we wanted to minimize the function $P(k)$ over the integers, not over
the reals. However, if we choose the integer $k$ so that $\frac{6}{6+k}$
is close to $e^{-1}$, one can see that the value of $P(k)$ is close to
$P(e^{-1})$. The minimum is in fact achieved at $k=10$. Thus $P(k)$ can be as small 
as $e^{-cN}$, where $c$ is some constant `close' to $\frac{1}{6e}$: 
It is in fact $c=\frac{1}{16} (\log 8 - \log 3) \approx 0,0613018$.

\section*{Appendix B: Library and Tables}

Table~\ref{tbl:discrete_morse_spectra} lists computational results
for the examples of Section~\ref{sec:computational_results}.

Of each example we present the discrete Morse spectrum we experimentally observed in a certain number of runs (usually 10000, when not otherwise stated; sometimes we did fewer runs for reasons concerning either excessive computation time or excessive variance of the spectrum). 

Let $c_{\approx}$ and $c^N_{\approx}$ be the average numbers of critical cells for the vectors 
in the approximated discrete Morse spectrum and the approximated normalized discrete Morse spectrum,
respectively.  The longer we run \textsc{Random Discrete Morse}, the better the approximation of  $c_{\sigma}$ by  $c_{\approx}$
and of $c^N_{\sigma}$ by  $c^N_{\approx}$ will get --- and possibly optimal discrete Morse vectors will show up.

In Table~\ref{tbl:discrete_morse_spectra}, optimal discrete Morse vectors are highlighted in bold. We wrote an output vector in italics if it is the best we could find with our algorithm and we do not know if it is indeed the optimal or not. 

For Table~\ref{tbl:lex_rev_lex}, we replaced the random choices in our algorithm 
with a deterministic lexicographic or reverse lexicographic choice.
The labeling of the vertices of course now plays a role; see~\cite{AdiprasitoBenedettiLutz2013pre}
for a discussion of a randomized version (by randomly renumbering vertices first) of \texttt{lex} and \texttt{rev\_lex}. 

All computations were run on a cluster of 2.6 GHz processors.

\pagebreak

{\small
\defaultaddspace=.1em

\setlength{\LTleft}{0pt}
\setlength{\LTright}{0pt}
\begin{longtable}{@{}l@{\extracolsep{\fill}}l@{\extracolsep{2pt}}r@{\extracolsep{\fill}}l@{}}
\caption{\protect\parbox[t]{15cm}{Library of triangulations and discrete Morse spectra.}}\label{tbl:discrete_morse_spectra}
\\\toprule
 \addlinespace
 \addlinespace
 \addlinespace
 \addlinespace
 Name of example/Homology/                      & \multicolumn{2}{@{}l@{}}{Distribution of obtained} & Time for Hasse diagram/   \\
 $f$-vector/$c^{N}_{\approx}$ & \multicolumn{2}{@{}l@{}}{discrete Morse vectors}   & Time per round             \\ 
                                                   & \multicolumn{2}{@{}l@{}}{in 10000 rounds}         & (in Hour:Min:Sec.Frac) \\ \midrule
\endfirsthead
\caption{\protect\parbox[t]{15cm}{Library of triangulations and discrete Morse spectra (continued).}}
\\\toprule
 \addlinespace
 \addlinespace
 \addlinespace
 \addlinespace
 Name of example/Homology/                      & \multicolumn{2}{@{}l@{}}{Distribution of obtained} & Time for Hasse diagram/   \\
 $f$-vector/$c^{N}_{\approx}$ & \multicolumn{2}{@{}l@{}}{discrete Morse vectors}   & Time per round             \\ 
                                                   & \multicolumn{2}{@{}l@{}}{in 10000 rounds}         & (in Hour:Min:Sec.Frac) \\ \midrule
\endhead
\bottomrule
\endfoot
  \texttt{dunce\_hat}                           & \textbf{(1,1,1)}:       & 10000 & 0.004 \\[-1mm]
  $(\mathbb{Z},0,0)$                            &                         &       & 0.00024 \\[-1.1mm]
  $(8,24,17)$                                   &                         &       & \\[-1.1mm]
  $3.0000$                                      &                         &       & \\[-.25mm]
 \addlinespace
 \addlinespace
  \texttt{d2n12g6}                              & \textbf{(1,12,1)}:      &  9722 & 0.004 \\[-1mm]
  $(\mathbb{Z},\mathbb{Z}^{12},\mathbb{Z})$     & $(2,13,1)$:             &   277 & 0.00076 \\[-1.1mm] 
  $(12,66,44)$                                  & $(3,14,1)$:             &     1 & \\[-1.1mm] 
  $14.0000$                                     &                         &       & \\
\end{longtable}

}


{\small
\defaultaddspace=.1em

\setlength{\LTleft}{0pt}
\setlength{\LTright}{0pt}
\begin{longtable}{@{}l@{\extracolsep{\fill}}l@{\extracolsep{\fill}}l@{}}
\caption{\protect\parbox[t]{15cm}{Discrete Morse vectors with \texttt{lex} and \texttt{rev\_lex} heuristics.}}\label{tbl:lex_rev_lex}
\\[-.7mm]\toprule
 \addlinespace
 \addlinespace
 \addlinespace
 \addlinespace
 Name of example  &    lex     & rev\_lex      \\[-.7mm] \midrule
\endfirsthead
\caption{\protect\parbox[t]{15cm}{Discrete Morse vectors with \texttt{lex} and \texttt{rev\_lex} heuristics (continued).}}
\\[-.7mm]\toprule
 \addlinespace
 \addlinespace
 \addlinespace
 \addlinespace
 Name of example  &    lex     & rev\_lex      \\[-.7mm] \midrule
\endhead
\bottomrule
\endfoot
 \addlinespace
 \addlinespace
 \addlinespace
 \addlinespace
  \texttt{dunce\_hat}                                                        & \textbf{(1,1,1)}             & \textbf{(1,1,1)}             \\[-0.5mm]
  \texttt{d2n12g6}                                                           & \textbf{(1,12,1)}            & \textbf{(1,12,1)}            \\[-0.5mm]
  \texttt{regular\_2\_21\_23\_1}                                             & \textbf{(1,30,1)}            & \textbf{(1,30,1)}            \\[-0.5mm]
  \texttt{rand2\_n25\_p0.328}                                                & \textbf{(1,0,475)}           & \textbf{(1,0,475)}           \\[-0.5mm]
  \texttt{dunce\_hat\_in\_3\_ball}                                           & \textbf{(1,0,0,0)}           & \textbf{(1,0,0,0)}           \\[-0.5mm]
  \texttt{Barnette\_sphere}                                                  & \textbf{(1,0,0,1)}           & \textbf{(1,0,0,1)}           \\[-0.5mm]
  \texttt{B\_3\_9\_18} (non-shellable ball)                                  & \textbf{(1,0,0,0)}           & \textbf{(1,0,0,0)}           \\[-0.5mm]
  \texttt{trefoil\_arc}                                                      & $(1,2,2,0)$                  & \textbf{(1,0,0,0)}           \\
\end{longtable}

}

\pagebreak

{\small
\defaultaddspace=.1em

\setlength{\LTleft}{0pt}
\setlength{\LTright}{0pt}
\begin{longtable}{@{}l@{\extracolsep{\fill}}r@{\extracolsep{\fill}}r@{\extracolsep{\fill}}r@{\extracolsep{\fill}}r@{\extracolsep{\fill}}r@{\extracolsep{\fill}}r@{}}
\caption{\protect\parbox[t]{15cm}{Simplified presentations of
    fundamental groups with GAP.}}\label{tbl:gen_fund_groups}
\\[-.7mm]\toprule
 \addlinespace
 \addlinespace
 \addlinespace
 \addlinespace
 Name of example  &   Ge.\ &  Re.\ &  SGe.\ & SRe.\ & F.~Gr.\ & Time  \\[-.7mm] \midrule
\endfirsthead
\caption{\protect\parbox[t]{15cm}{Simplified presentations of
    fundamental groups with GAP (continued).}}
\\[-.7mm]\toprule
 \addlinespace
 \addlinespace
 \addlinespace
 \addlinespace
 Name of example  &   Ge.\ &  Re.\ &  SGe.\ & SRe.\ & F.~Gr.\ & Time  \\[-.7mm] \midrule
\endhead
\bottomrule
\endfoot
 \addlinespace
 \addlinespace
 \addlinespace
 \addlinespace
  \texttt{dunce\_hat}                                                        &    17 &     17 &  0 &  0 &              0 &       0.036 \\[-0.5mm]
  \texttt{d2n12g6}                                                           &    55 &     44 & 12 &  1 &        12 gen. &       0.132 \\[-0.5mm]
  \texttt{regular\_2\_21\_23\_1}                                             &   127 &     98 & 30 &  1 &        30 gen. &       0.304 \\[-0.5mm]
  \texttt{rand2\_n25\_p0.328}                                                &   276 &    751 &  0 &  0 &              0 &       0.876 \\[-0.5mm]
  \texttt{dunce\_hat\_in\_3\_ball}                                           &    18 &     30 &  0 &  0 &              0 &       0.044 \\[-0.5mm]
  \texttt{Barnette\_sphere}                                                  &    20 &     38 &  0 &  0 &              0 &       0.048 \\[-0.5mm]
  \texttt{B\_3\_9\_18} (non-shellable ball)                                  &    25 &     43 &  0 &  0 &              0 &       0.048 \\[-0.5mm]
  \texttt{trefoil\_arc}                                                      &    47 &     85 &  0 &  0 &              0 &       0.092 \\
\end{longtable}

}

\begin{table}
\small\centering
\defaultaddspace=0.15em
\caption{Distribution of fundamental groups for random $2$-complexes
  with $25$ vertices (100~runs for each $p$).}\label{tbl:rand_2compl_n25}
\begin{tabular*}{\linewidth}{@{\extracolsep{\fill}}l@{\hspace{9mm}}r@{\hspace{2mm}}r@{\hspace{2mm}}r@{\hspace{2mm}}r@{\hspace{2mm}}r@{\hspace{2mm}}r@{\hspace{2mm}}r@{\hspace{2mm}}r@{\hspace{2mm}}r@{\hspace{2mm}}r@{\hspace{2mm}}r@{\hspace{2mm}}r@{\hspace{2mm}}r@{}}
\\\toprule
 \addlinespace
 \addlinespace
  $p$ & $0$ & $F(1)$ & $F(2)$ & $F(3)$ & $F(4)$ & $F(5)$ & $F(6)$ & $F(7)$ & $F(8)$ & $F(9)$ & $F(10)$ & $F(\geq 11)$ & `non-free' \\ \midrule
 \addlinespace
 \addlinespace
 0.40 & 100 &      0 &      0 &      0 &      0 &      0 &      0 &      0 &      0 &      0 &       0 &                  0 &        0 \\[-0.5mm]
 0.39 & 100 &      0 &      0 &      0 &      0 &      0 &      0 &      0 &      0 &      0 &       0 &                  0 &        0 \\[-0.5mm]
 0.38 &  99 &      1 &      0 &      0 &      0 &      0 &      0 &      0 &      0 &      0 &       0 &                  0 &        0 \\[-0.5mm]
 0.37 & 100 &      0 &      0 &      0 &      0 &      0 &      0 &      0 &      0 &      0 &       0 &                  0 &        0 \\[-0.5mm]
 0.36 &  98 &      2 &      0 &      0 &      0 &      0 &      0 &      0 &      0 &      0 &       0 &                  0 &        0 \\[-0.5mm]
 0.35 &  96 &      3 &      1 &      0 &      0 &      0 &      0 &      0 &      0 &      0 &       0 &                  0 &        0 \\
\bottomrule
\end{tabular*}
\end{table}

\begin{table}
\small\centering
\defaultaddspace=0.15em
\caption{Distribution of fundamental groups for random $2$-complexes
  with $50$ vertices (100~runs for each $p$).}\label{tbl:rand_2compl_n50}
\begin{tabular*}{\linewidth}{@{\extracolsep{\fill}}l@{\hspace{9mm}}r@{\hspace{2mm}}r@{\hspace{2mm}}r@{\hspace{2mm}}r@{\hspace{2mm}}r@{\hspace{2mm}}r@{\hspace{2mm}}r@{\hspace{2mm}}r@{\hspace{2mm}}r@{\hspace{2mm}}r@{\hspace{2mm}}r@{\hspace{2mm}}r@{\hspace{2mm}}r@{}}
\\\toprule
 \addlinespace
 \addlinespace
  $p$ &   0 & $F(1)$ & $F(2)$ & $F(3)$ & $F(4)$ & $F(5)$ & $F(6)$ & $F(7)$ & $F(8)$ & $F(9)$ & $F(10)$ & $F(\geq 11)$ & `non-free' \\\midrule
 \addlinespace
 \addlinespace
 0.25 & 100 &      0 &      0 &      0 &      0 &      0 &      0 &      0 &      0 &      0 &       0 &                  0 &        0 \\[-0.5mm]
 0.24 & 100 &      0 &      0 &      0 &      0 &      0 &      0 &      0 &      0 &      0 &       0 &                  0 &        0 \\[-0.5mm]
 0.23 &  99 &      1 &      0 &      0 &      0 &      0 &      0 &      0 &      0 &      0 &       0 &                  0 &        0 \\[-0.5mm]
 0.22 &  99 &      1 &      0 &      0 &      0 &      0 &      0 &      0 &      0 &      0 &       0 &                  0 &        0 \\[-0.5mm]
 0.21 &  97 &      3 &      0 &      0 &      0 &      0 &      0 &      0 &      0 &      0 &       0 &                  0 &        0 \\[-0.5mm]
 0.20 &  98 &      2 &      0 &      0 &      0 &      0 &      0 &      0 &      0 &      0 &       0 &                  0 &        0 \\
\bottomrule
\end{tabular*}
\end{table}

\pagebreak

\bibliography{.}

\bigskip
\bigskip
\smallskip

\small

\noindent
Bruno Benedetti \\
Institut f\"ur Informatik \\
Freie Universit\"at Berlin\\
Takustr.\ 9  \\
14195 Berlin, Germany \\
{\tt bruno@zedat.fu-berlin.de}

\vspace{.9cm}

\noindent
Frank H. Lutz\\
Institut for Matematiske Fag \\
K{\o}benhavns Universitet \\
Universitetsparken 5 \\
2100 K{\o}benhavn {\O}, Denmark\\
{\tt lutz@math.ku.dk}

\end{document}